\documentclass[fleqn,usenatbib]{mnras}

\usepackage{newtxtext,newtxmath}
\usepackage{CJKutf8}
\usepackage{lscape}

\usepackage[T1]{fontenc}

\DeclareRobustCommand{\VAN}[3]{#2}
\let\VANthebibliography\thebibliography
\def\thebibliography{\DeclareRobustCommand{\VAN}[3]{##3}\VANthebibliography}

\usepackage{graphicx}	% Including figure files
\usepackage{amsmath}	% Advanced maths commands
\usepackage{multirow}
\usepackage{lscape}
\usepackage{makecell}

\title[Two late-T dwarfs at kiloparsec distances]{Two late-T dwarfs at kiloparsec distances revealed by  \textit{JWST} UNCOVER survey}

\author[D. H. Li et al.]{D. H. Li (\begin{CJK*}{UTF8}{gbsn}李东恒\end{CJK*}),$^{1,2,3}$ Z. H. Zhang (\begin{CJK*}{UTF8}{gbsn}张曾华\end{CJK*}),$^{1,2}$\thanks{E-mail: zz@nju.edu.cn} H. H. Peng (\begin{CJK*}{UTF8}{gbsn}彭浩晖\end{CJK*}),$^{1,2}$  M. C. G\'alvez-Ortiz,$^{4}$ 
\newauthor
S. Y. Zhou (\begin{CJK*}{UTF8}{gbsn}周思琰\end{CJK*})$^{1,2,3}$ and H. R. A. Jones$^{5}$
\\
$^{1}$School of Astronomy and Space Science, Nanjing University, 163 Xianlin Avenue, Nanjing 210023, China \\
$^{2}$Key Laboratory of Modern Astronomy and Astrophysics, Nanjing University, Ministry of Education, Nanjing 210023, China \\
$^{3}$School of Physical Science and Technology, ShanghaiTech University, 100 Haike Road, Shanghai 201210, China \\
$^{4}$Centro de Astrobiolog{\'i}a (CAB), CSIC-INTA, Camino Bajo del Castillo s/n, E-28692, Villanueva de la Ca{\~n}ada, Madrid, Spain \\
$^{5}$Centre for Astrophysics Research, University of Hertfordshire, Hatfield, Hertfordshire AL10 9AB, UK 
}

\date{Accepted 2026 January 30. Received 2026 January 30; in original form 2025 July 17}

\pubyear{\the\year{}}

\begin{document}
\label{firstpage}
\pagerange{\pageref{firstpage}--\pageref{lastpage}}
\maketitle

\begin{abstract}
We conducted a search for brown dwarf candidates in a \textit{James Webb Space Telescope} deep field around A2744 to investigate the space density of these objects at kiloparsec distances. Our methodology employed an initial selection based on photometric colours, followed by spectral energy distribution fitting to both stellar atmospheric models and high-redshift galaxy templates. This approach yielded two robust T dwarf candidates and one possible L subdwarf candidate. The T dwarfs have estimated Galactic heights of 0.43 and 0.86~kpc, likely residing near the outer edges of the Galactic thin and thick discs, respectively.
We measure a T dwarf surface number density of 0.094~arcmin$^{-2}$ in the UNCOVER field, lower than previous predictions but consistent at the order-of-magnitude level. 
We also provide space number density estimates for T5–T8.9 dwarfs across different effective temperature and spectral type bins, finding that T5–T7 dwarfs out to 2~kpc have significantly lower densities than their solar neighbourhood counterparts, whilst T8 dwarfs within the thick disc exhibit densities comparable to local values. 
Our analysis demonstrates that broad-band near- to mid-infrared photometry provides high sensitivity to late-T dwarfs but is relatively less sensitive to L and early-T dwarfs. Spectroscopy is typically required to distinguish photometric candidates of L dwarfs, early-T subdwarfs, and high-redshift galaxies in \textit{JWST} deep fields.
This study demonstrates the potential for expanding our understanding of brown dwarf distributions and characteristics at unprecedented distances, offering new insights into substellar populations beyond the solar neighbourhood.

\end{abstract}


%%%%%%%%%%%%%%%%%%%%%%%%%%%%%%%%%%%%%%%%%%%%%%%%%%

%%%%%%%%%%%%%%%%% BODY OF PAPER %%%%%%%%%%%%%%%%%%

\section{Introduction}
Brown dwarfs (BDs) are substellar objects that bridge the mass gap between the most massive gaseous planets and the lowest-mass stars, typically ranging from 0.01 to 0.08 M$_{\odot}$ \citep{Kuma63,Haya63,prime6,Marl21,chab23}. Unlike main-sequence stars, BDs lack sufficient mass to sustain steady hydrogen fusion in their cores and instead rely primarily on their initial thermal energy, cooling continuously over time.

Three spectral classes have been established to classify BDs according to their effective temperature ($T_{\rm eff}$) ranges and spectral characteristics: L dwarfs (2300 K $\gtrsim T_{\rm eff} \gtrsim 1300$ K; \citealt{kirk99,mart99}), T dwarfs (1300 K $\gtrsim T_{\rm eff} \gtrsim$ 500 K; \citealt{burg02}), and Y dwarfs ($T_{\rm eff} \lesssim$ 500 K; \citealt{cush11}). The discovery and characterization of BDs are crucial for constraining the initial mass function and advancing our understanding of ultracool exoplanetary atmospheres.

Statistically, the majority of BDs in the Milky Way have spectral types later than T5 \citep{reyl21}. These cool BDs have effective temperatures ($T_{\rm eff}$) below $\sim$1000 K and emit most of their flux in the near-infrared (NIR) and mid-infrared, making these wavelengths optimal for their detection. However, due to the limited depths of large-area surveys, most known T5+ dwarfs are located within 20 pc of the Sun \citep[e.g.,][]{Kirk21}.

The launch of the \textit{James Webb Space Telescope} (\textit{JWST}) has ushered in a new era of BD research. Its Near Infrared Camera \citep[NIRCam;][]{nircam} provides photometric data from 0.6 to 5.0 $\mu$m with unprecedented depth, reaching AB magnitudes of $\sim$30. This capability enables the detection of distant BDs beyond the Galactic plane. Several studies have predicted the number of BDs observable in \textit{JWST} deep surveys \citep{Ryan16,agan22b,Case23}, and multiple BD candidates have been identified in \textit{JWST} deep multi-band imaging surveys \citep{Glaz23,Noni23,Wang23,Hain24,hain25,Holw24,chen25,}.  its Near Infrared Spectrograph \citep[NIRSpec;][]{nirspec}, several of these BD candidates in \textit{JWST} deep fields have been spectroscopically confirmed \citep{Lang23,Bur24,Hainl24,tu25b,tu25,morr25}.

We aimed to identify new BD candidates through spectral energy distribution (SED) model fitting using data from the \textit{JWST} Cycle 1 treasury programme Ultradeep NIRSpec and NIRCam ObserVations before the Epoch of Reionization \citep[UNCOVER;][]{Bezan24} and the Cycle 2 MegaScience Survey \citep{Sues24}. In this paper, we present two robust late T dwarf candidates and one possible L subdwarfs identified in the Abell 2744 field. Section \ref{data} describes our selection method for BD candidates. Section \ref{results} presents the properties of the new BD candidates. In Section \ref{dicuss}, we discuss the Galactic distribution of UNCOVER BDs. Finally, we summarize our findings in Section \ref{conclusion}.

\begin{figure*}
    \centering
    \includegraphics[width=0.97\textwidth]{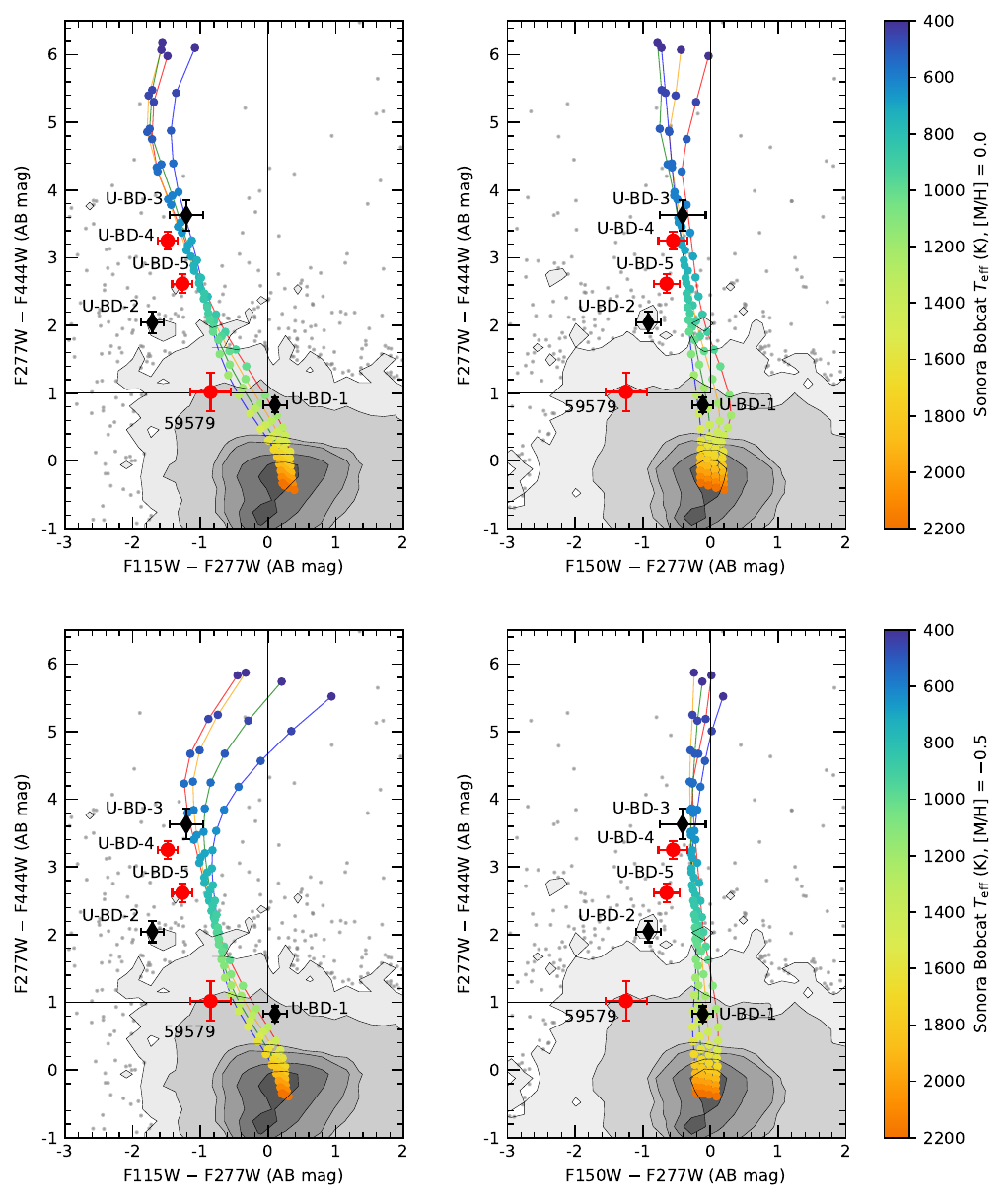}
    \caption{
    \textit{JWST} NIRCam colour–colour diagrams showing the three new BD candidates (red circles) and three known BDs (black diamonds) in the UNCOVER field. Grey points represent all sources in the UNCOVER catalogue, with black contour lines indicating the 25th, 50th, 68th, 75th, 95th, and 99th percentiles of the source density distribution. Solid black lines delineate our selection criteria defined in Equations~\ref{eq1}–\ref{eq2}. Coloured points show theoretical predictions from the Sonora atmospheric models \citep{Marl21} for $T_{\rm eff}$ ranging from 400 to 2200~K (in steps of 50~K for $T_{\rm eff} < 1000$~K and 100~K for $T_{\rm eff} > 1000$~K), with surface gravities of $4.0 \leq \log g \leq 5.5$ and metallicities of $\rm [M/H] = 0.0$ (upper panels) and $\rm [M/H] = -0.5$ (lower panels). The colour scale indicates temperature, transitioning from orange (hottest) to blue (coolest). Models with identical $\log g$ values are connected by lines: red ($\log g = 4.0$), orange ($\log g = 4.5$), green ($\log g = 5.0$), and blue ($\log g = 5.5$). Model magnitudes have been converted from the Vega to the AB photometric system using the following offsets: 0.97 (F115W), 1.14 (F150W), 1.53 (F277W), and 2.97 (F444W).
    }
    \label{fcolor}
\end{figure*}

\section{Brown dwarf candidate identification}
\label{data}
\subsection{UNCOVER and MegaScience surveys}
\label{usurvey}
The UNCOVER second data release (DR2; \citealt{Weav24}) features a `SUPER' catalogue with photometry derived from optimally selected colour apertures, improved star and artefact identification, and complete coverage of all available \textit{JWST} imaging over Abell~2744 (49~arcmin$^2$). The third data release (DR3; \citealt{Sues24}) combines UNCOVER and MegaScience data, expanding the filter set to include additional NIRCam and NIRISS medium bands and providing photometric redshifts derived using Prospector-$\beta$ \citep{Wan23}. All catalogues are based on F444W PSF-matched imaging, with bright cluster galaxies and intracluster light subtracted.

Object detection was performed on a noise-equalized long-wavelength image (F277W+F356W+F444W) \citep{Fuji25}. Noise equalization homogenizes the noise distribution across the image by applying weighted corrections, ensuring a consistent signal-to-noise ratio (SNR) detection threshold. This approach enhances the reliability of faint source detection and reduces spurious detections. The resulting catalogues provide photometry, photometric redshifts, and rest-frame fluxes derived using \textsc{eazy} \citep[Easy and Accurate Redshifts from Yale;][]{Bram08}.

\subsection{Initial selection by colours}
\label{colour}

Both models and observed spectra \citep{Marl21,beil24} demonstrate that T dwarfs exhibit suppressed flux in the F277W band due to water and methane absorption features, as well as collision-induced absorption from H$_2$–H$_2$ \citep[CIA H$_2$; e.g.,][]{bory97}. Consequently, T dwarfs display blue F115W$-$F277W and F150W$-$F277W colours, alongside red F277W$-$F444W colours, as reflected in their SEDs \citep[e.g.,][]{Wang23,Hain24}. Figure~\ref{fcolor} presents UNCOVER sources in F115W$-$F277W versus F277W$-$F444W and F150W$-$F277W versus F277W$-$F444W colour–colour spaces, together with theoretical colours of ultracool dwarfs from the Sonora Bobcat models \citep{Marl21}.

\begin{figure}
    \centering
    \includegraphics[width=0.99\linewidth]{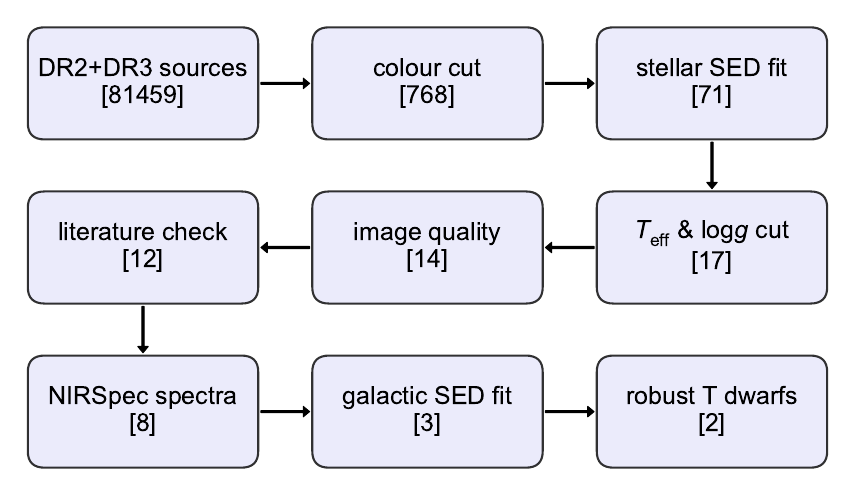}
    \caption{Flowchart of our selection process. The number of remaining candidates at each step is indicated.}
    \label{ffc}
\end{figure}

To identify cool BDs from UNCOVER based on their suppressed flux in the F277W band, we applied the following criteria:
\begin{eqnarray}
\label{eq1}
F277W - F444W \geq 1  \\
\label{eq2}
F115W - F277W \leq 0 ~{\rm OR}~ F150W - F277W \leq 0
\end{eqnarray}
We applied Equations~\ref{eq1}–\ref{eq2} (see Fig.\ref{fcolor}) to the UNCOVER DR2 and DR3 databases, prioritizing DR3 data for objects common to both releases. From a total of 81,459 sources, 768 preliminary candidates were selected. Fig.\ref{ffc} presents the flowchart of our selection process. \textit{JWST} magnitudes for selected BD candidates are listed in Table~\ref{tmag}.

\subsection{SED fitting to stellar atmospheric models}
\label{ssvosa}
\begin{figure}
    \centering
    \includegraphics[width=0.99\linewidth]{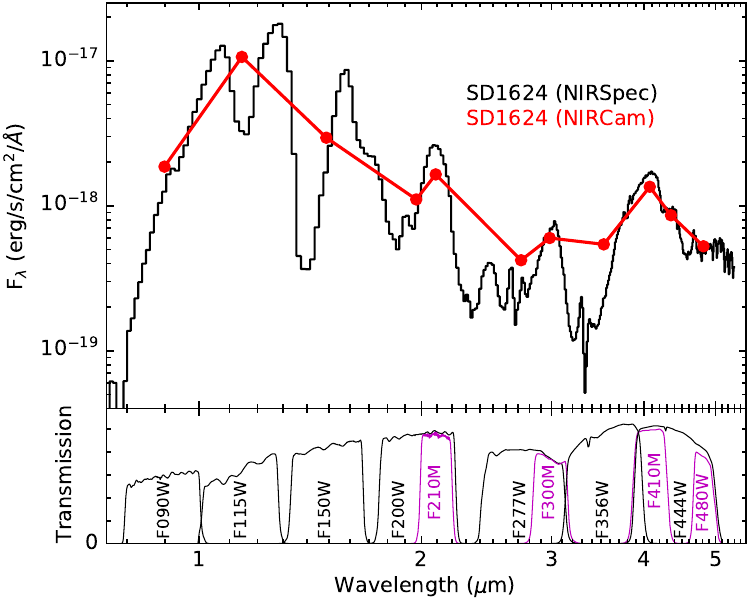} \\
    \includegraphics[width=0.99\linewidth]{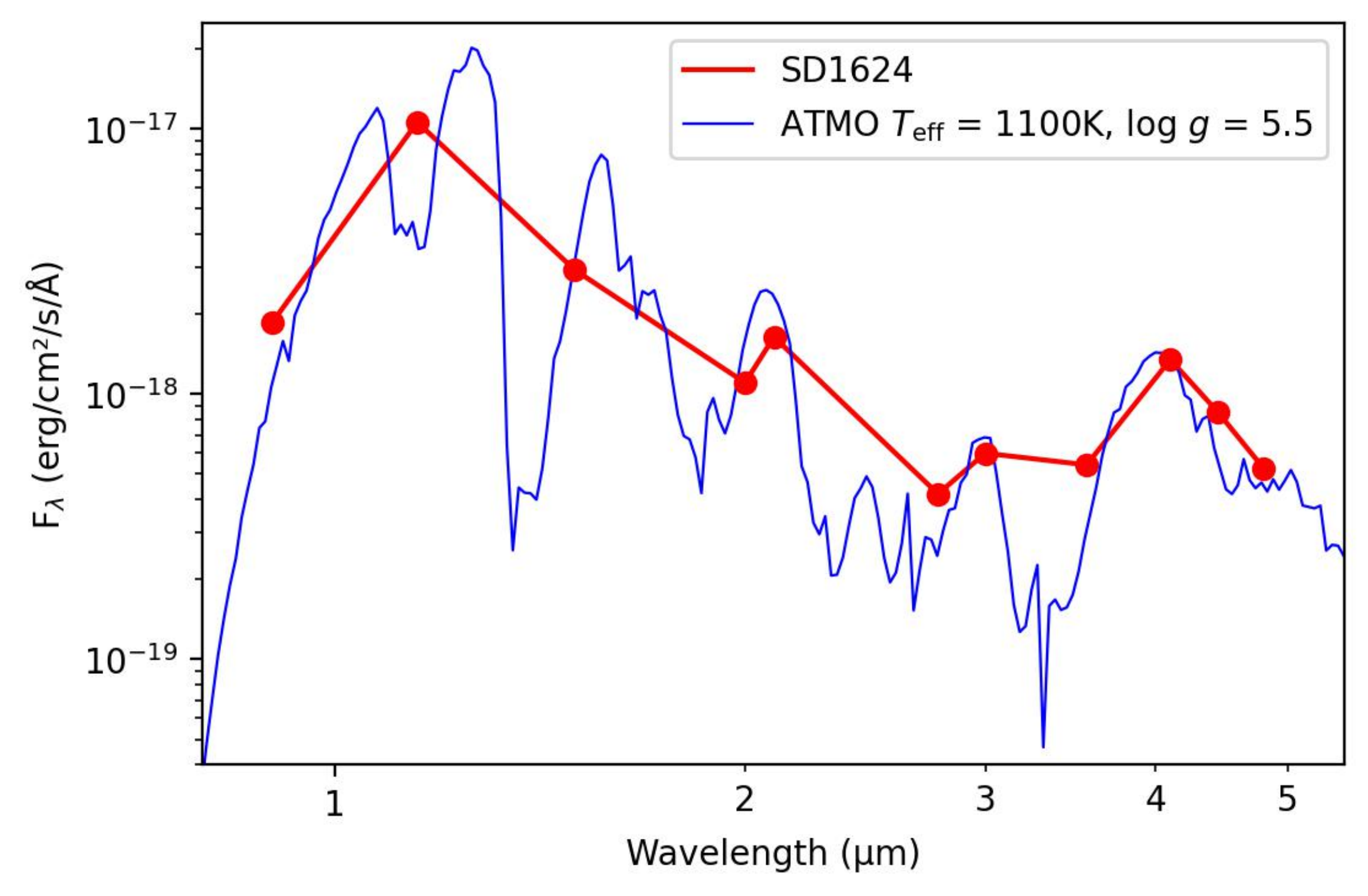} 
    \caption{
    Upper panel: \textit{JWST} NIRSpec/PRISM spectrum and NIRCam photometry of the T6 dwarf standard SD1624. The transmission profiles of NIRCam filters used in the UNCOVER survey are shown below the spectrum. Lower panel: best-fitting solar-metallicity ATMO model ($T_{\rm eff} = 1100$~K, $\log g = 5.5$) to the NIRCam photometry of SD1624 (red points).
    }
    \label{fsd1624}
\end{figure}

\begin{figure*}
    \includegraphics[width=0.495\linewidth]{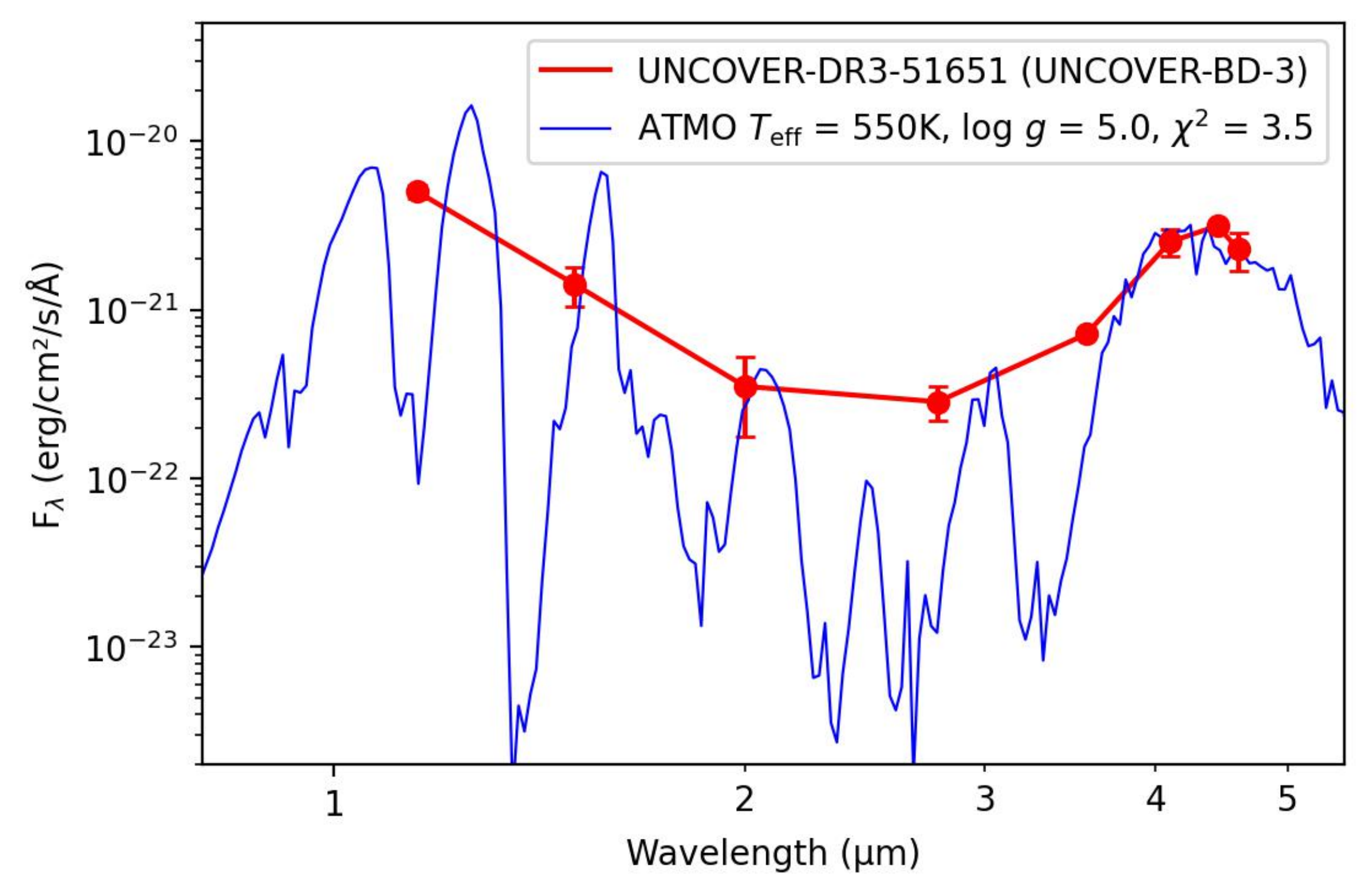}
    \includegraphics[width=0.495\linewidth]{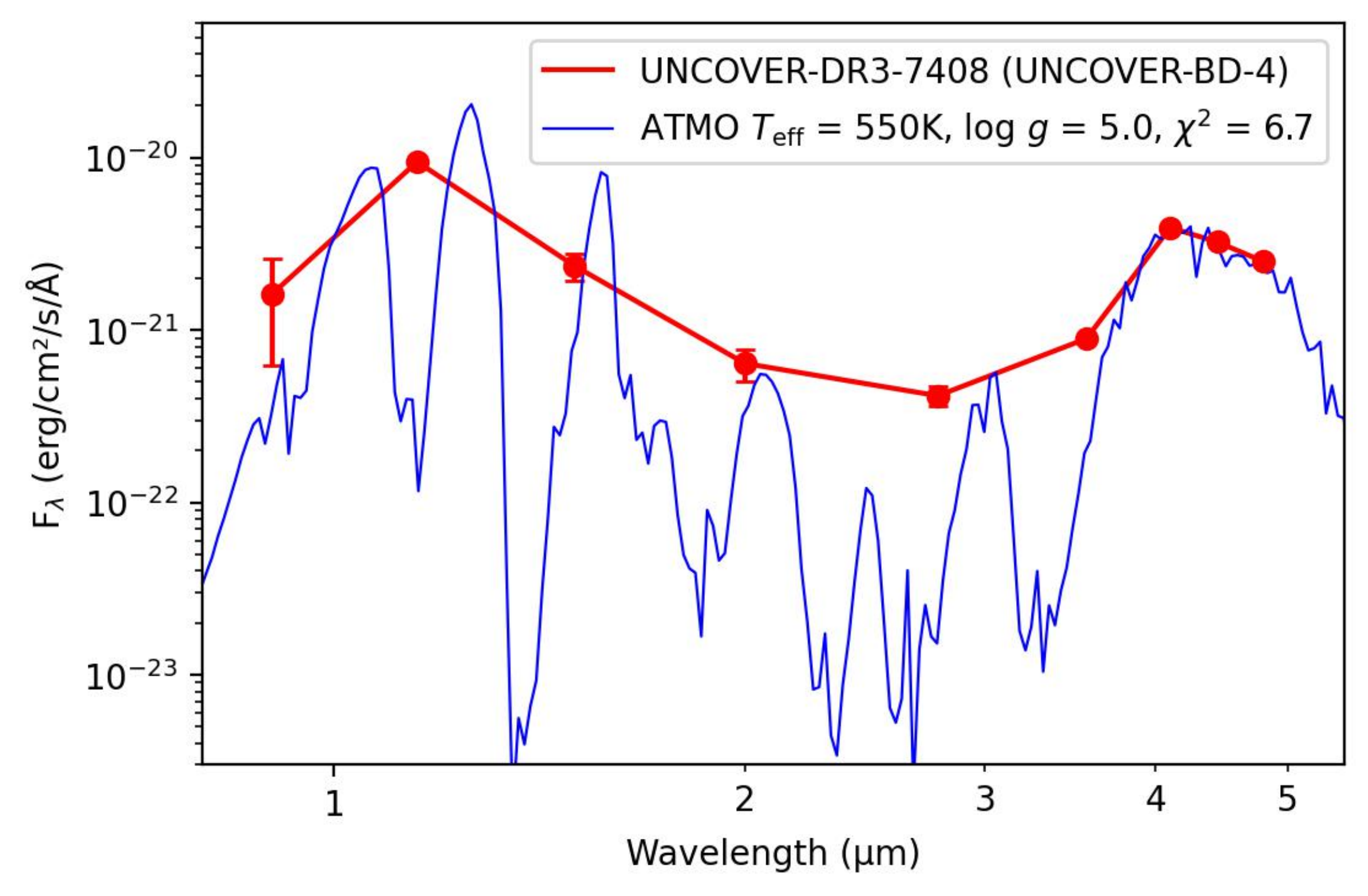} \\   
    \includegraphics[width=0.495\linewidth]{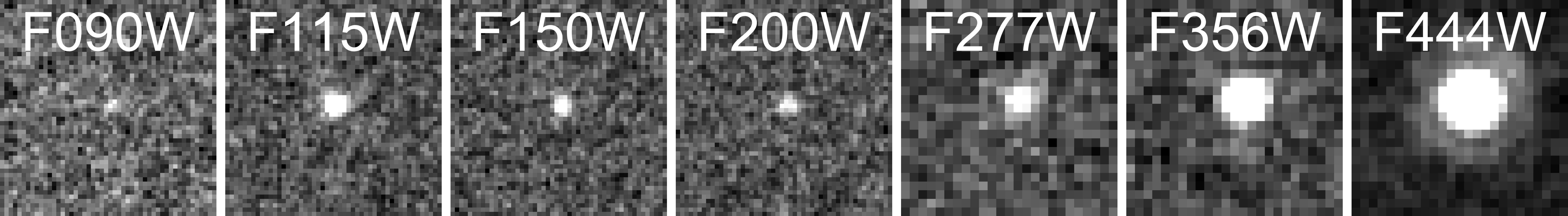}
    \includegraphics[width=0.495\linewidth]{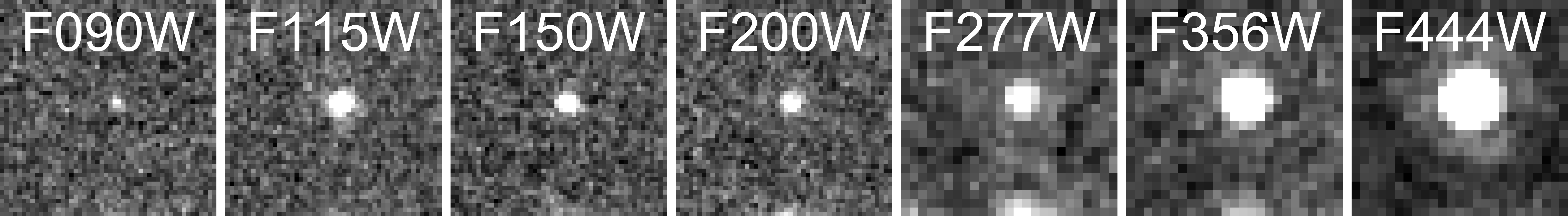}
    \includegraphics[width=0.495\linewidth]{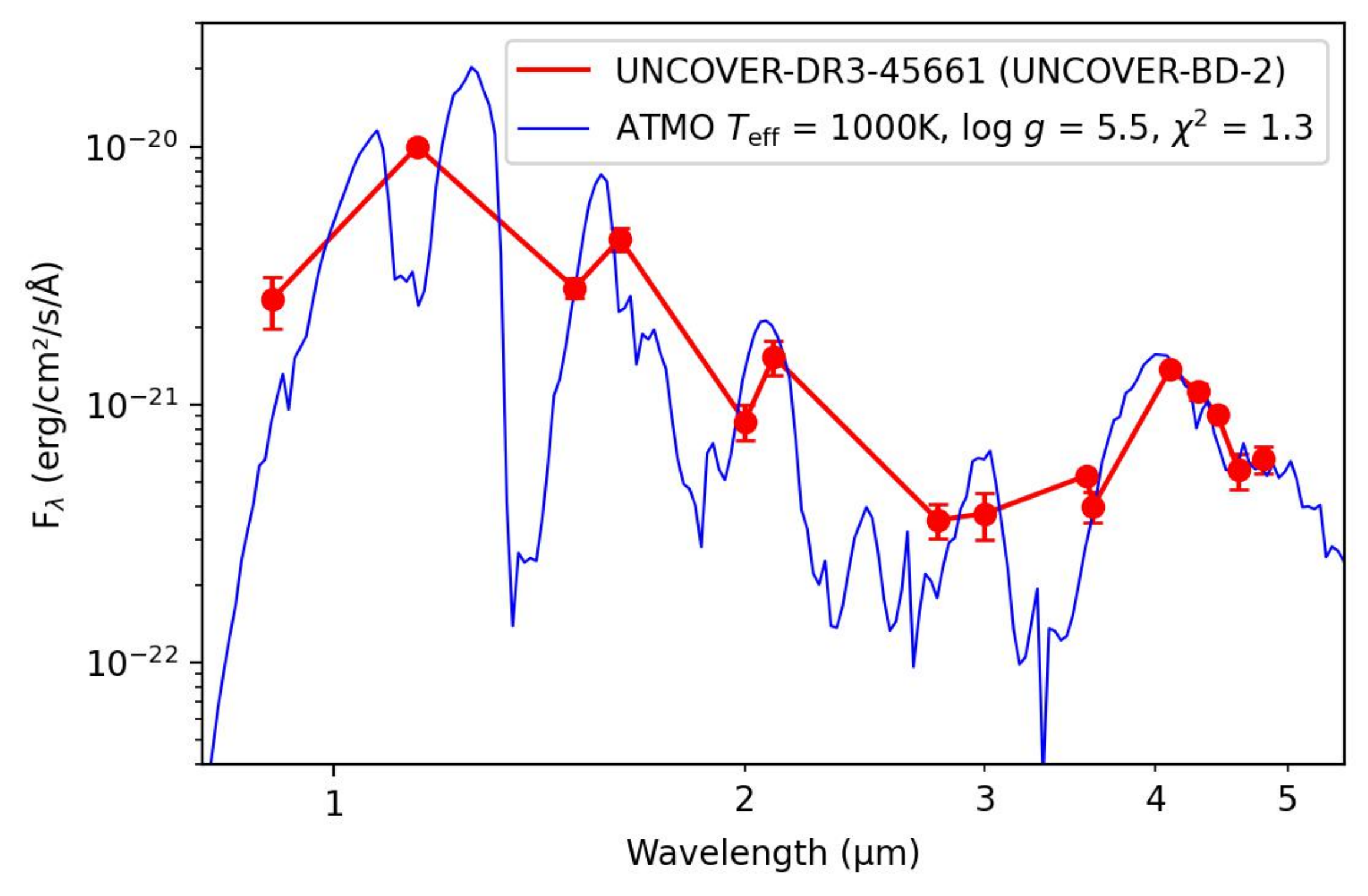} 
    \includegraphics[width=0.495\linewidth]{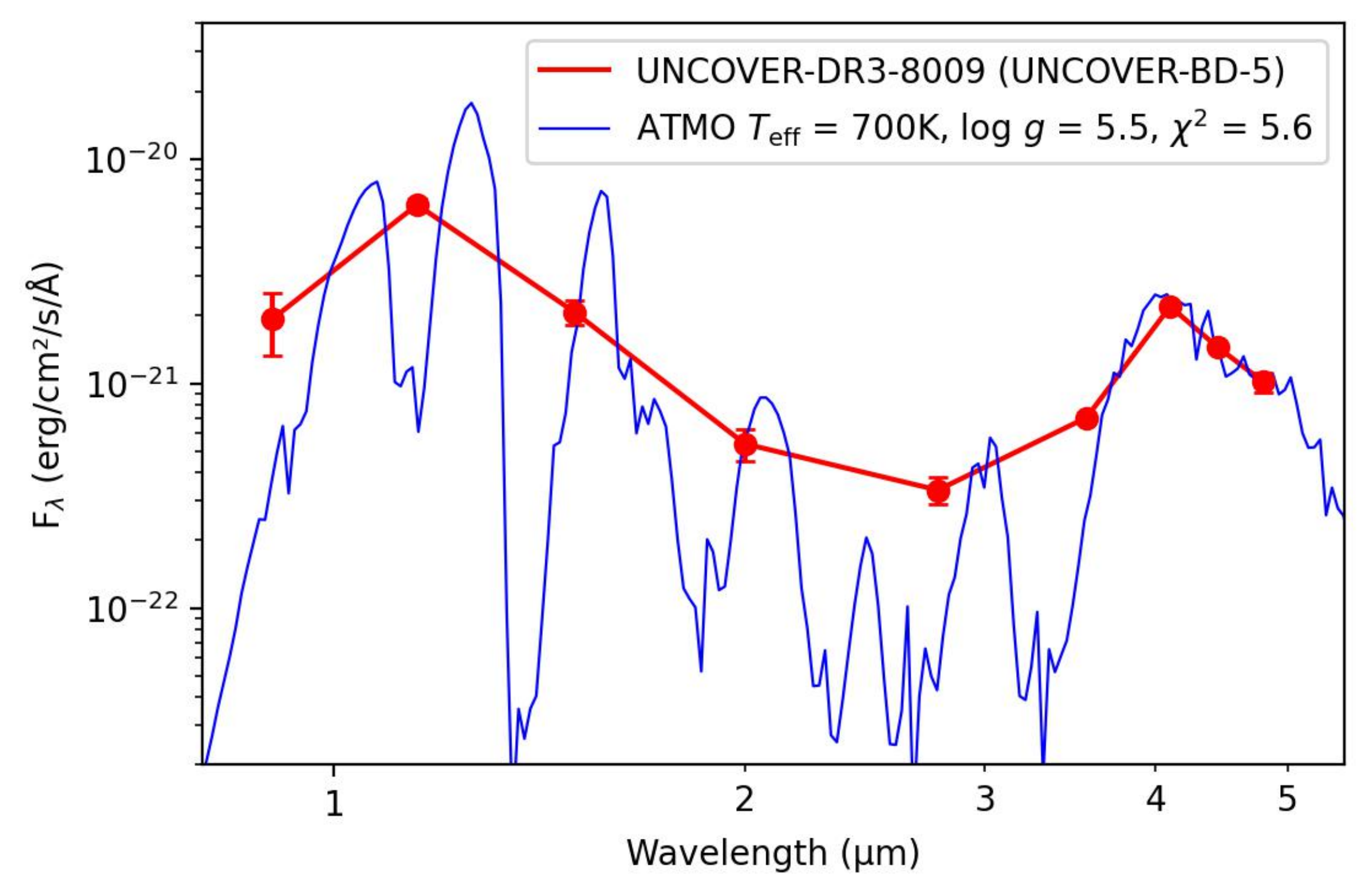} \\
    \includegraphics[width=0.495\linewidth]{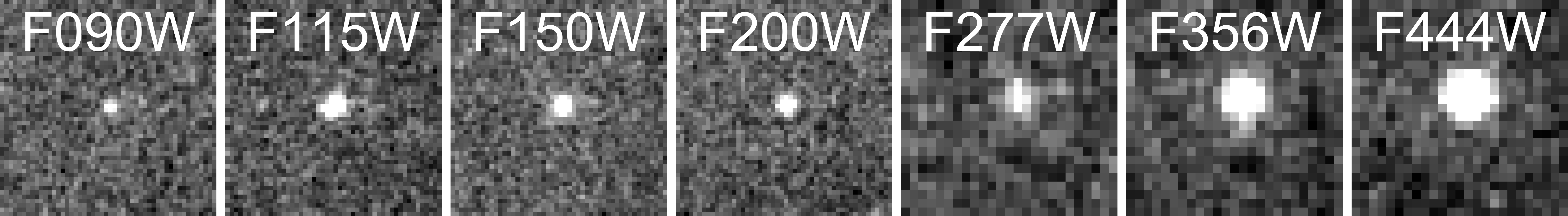} 
    \includegraphics[width=0.495\linewidth]{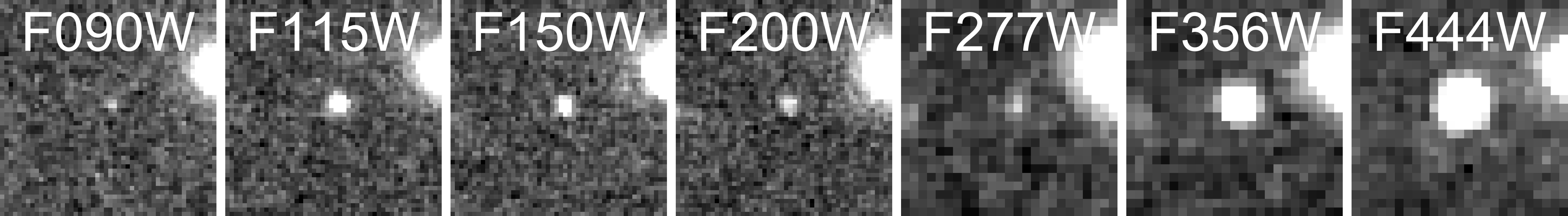} \\
    \includegraphics[width=0.495\linewidth]{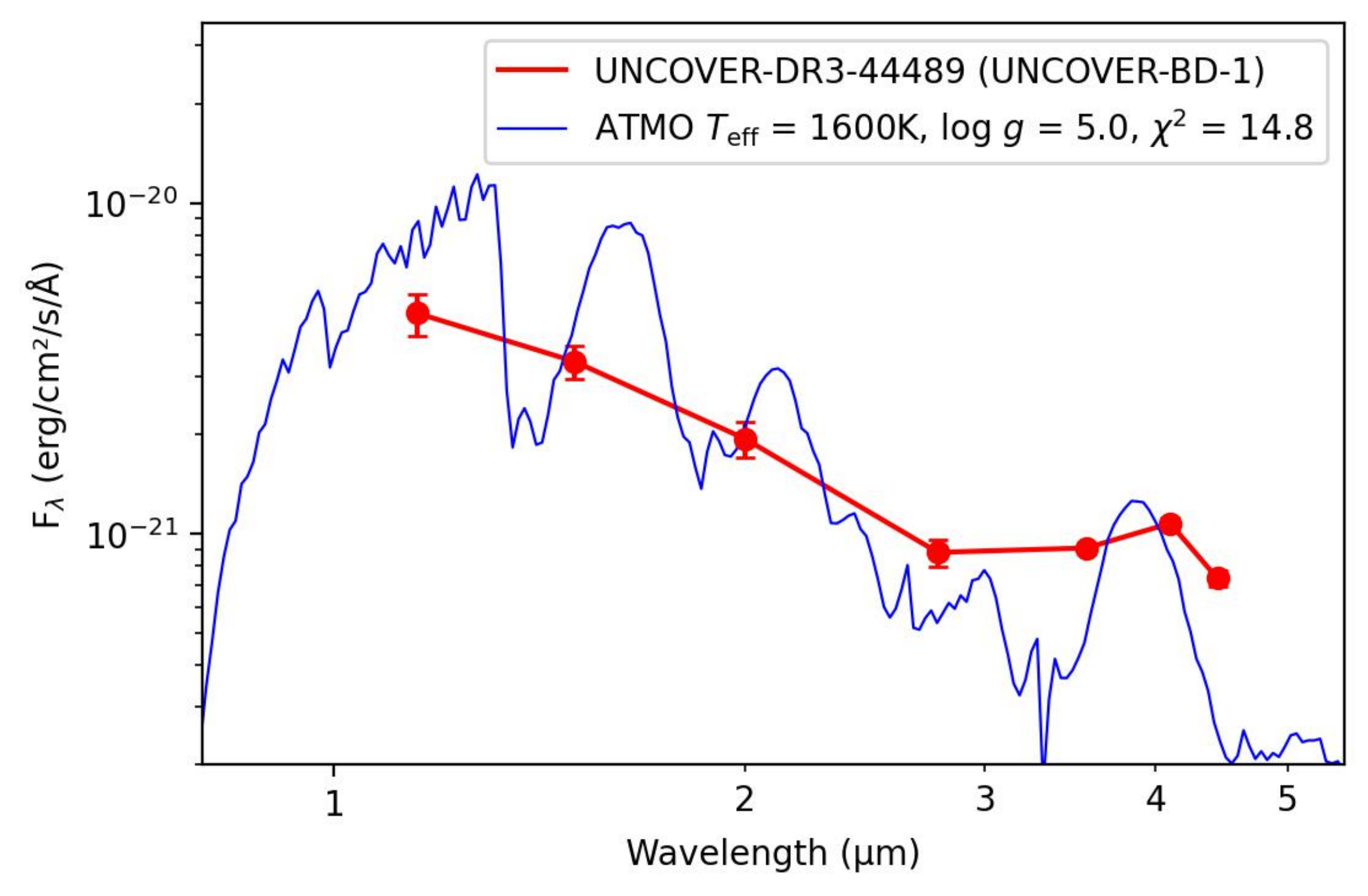} 
    \includegraphics[width=0.495\linewidth]{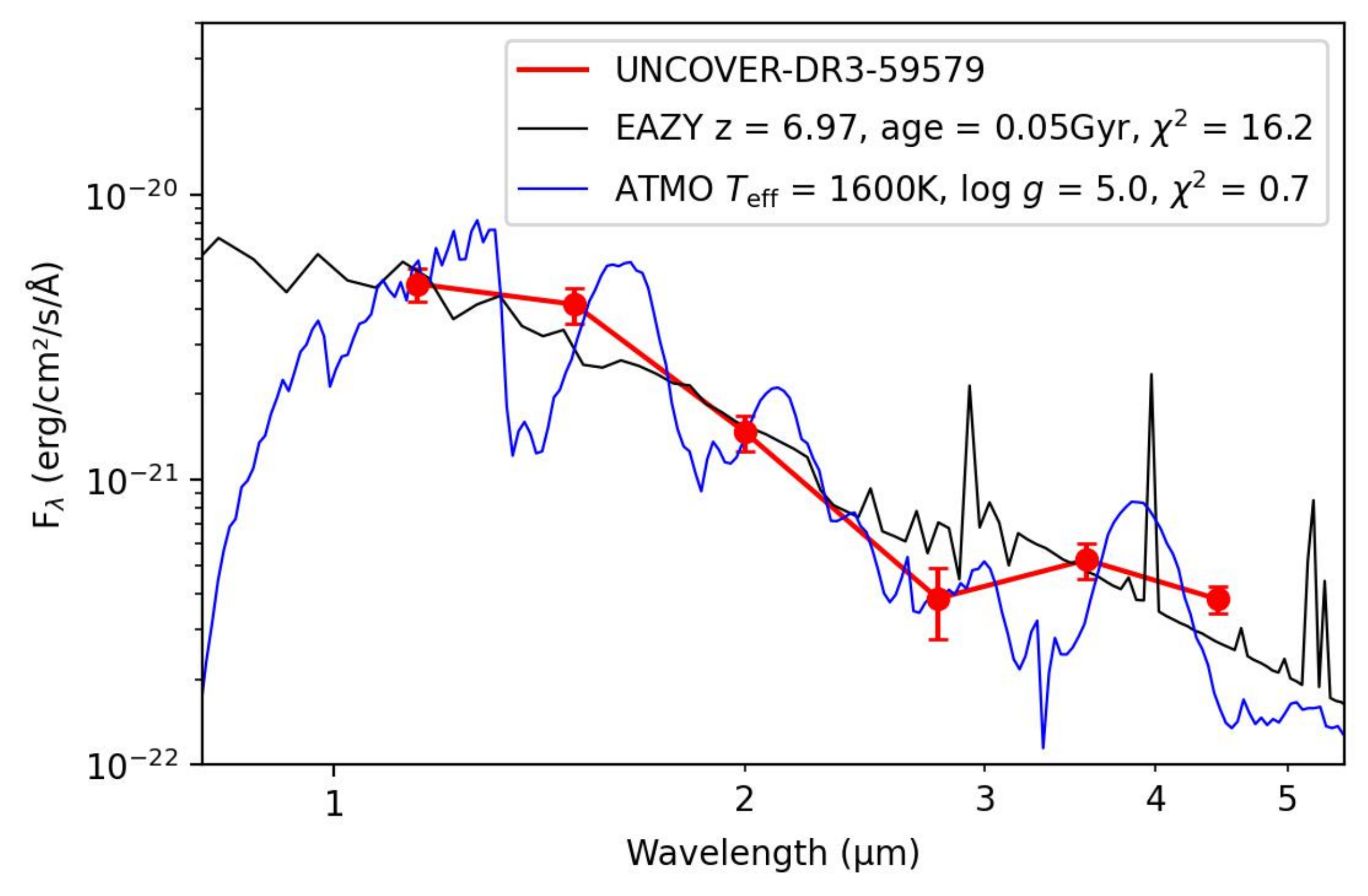} \\
    \includegraphics[width=0.495\linewidth]{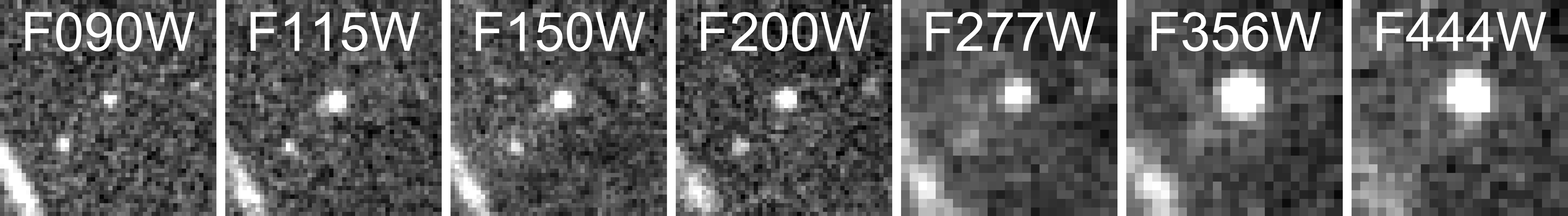} 
    \includegraphics[width=0.495\linewidth]{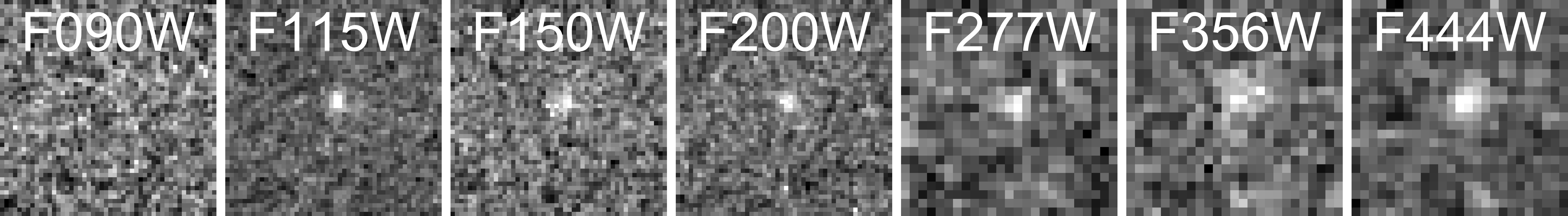} \\
    \caption{
    Best-fitting solar-metallicity ATMO models (blue spectra) to the observed SEDs (red points) of three known BDs (left-hand panels) and three new BD candidates (right-hand panels). \textit{JWST} NIRCam postage stamp images (F090W, F115W, F150W, F200W, F277W, F356W, F444W; $1\arcsec \times 1\arcsec$ field of view) are shown below each corresponding SED, with north up and east to the left. Interstellar extinction ($A_V$) was fixed at 0.0 for all fits. For UNCOVER-BD-1, NIRSpec spectral fitting \citep{Bur24} yields a best-fitting LOWZ model \citep{meis21} with $T_{\rm eff} = 1300$~K, $\log g = 5.0$, and $\rm [M/H] = -1.0$. The best-fitting galaxy template to UNCOVER-DR3-59579 is also overplotted.
    }
    \label{fsed}
\end{figure*}

Figure~\ref{fsd1624} (upper panel) presents the \textit{JWST} NIRSpec/PRISM spectrum and NIRCam SED of the T6 dwarf standard SDSS J162414.37+002915.6 \citep[SD1624;][]{beil24}. The photometric SED of SD1624 exhibits distinctive features, including flux suppression in the F277W band and peaks near 1.2 and 4.0~$\mu$m arising from water, methane, and CIA H$_2$ absorption. Notably, three or more photometric bands are required to reveal these peaks in the SED (e.g., F356W, F410M, and F444W for the 4.0~$\mu$m peak). These characteristics indicate that BD candidates can be identified based on their SEDs across $\sim$1--5~$\mu$m. Furthermore, Fig. \ref{fsd1624} (lower panel) demonstrates that the best-fitting ATMO model spectrum \citep{Phil20} to the NIRCam SED of SD1624 closely reproduces its NIRSpec spectrum (upper panel). Fig. \ref{fsed} (middle-left panel) shows the NIRCam SED of another T6 dwarf, UNCOVER-BD-2 \citep{Bur24}, along with its best-fitting ATMO model ($T{\rm eff} = 1000$~K, $\log g = 5.5$, $\rm [M/H] = 0.0$), which closely matches its best-fitting LOWZ model \citep{meis21} derived from its NIRSpec spectrum ($T_{\rm eff} = 1000$~K, $\log g = 5.25$, $\rm [M/H] = 0.0$). This concordance between the model fits and observed spectra validates the efficacy of using SED features for T dwarf identification and characterization.

To determine the atmospheric properties of the 768 colour-selected BD candidates, we constructed their SEDs using photometric data from UNCOVER and fitted their observed SEDs to various atmospheric models using the Virtual Observatory SED Analyzer \citep[VOSA;][]{Bayo08,Rodr24}. Our SED fitting process employed different sets of atmospheric models available through VOSA: BT-Settl \citep{Alla12,Aspl09,Caff11}, ATMO \citep{Phil20}, \citet{More12,More14}, and \citet{Saum12}. BT-Settl models provide three parameters: $T_{\rm eff}$, surface gravity ($\log g$), and metallicity [M/H], while the other models are limited to solar metallicity. We selected only sources with more than three photometric data points and SNR $\geq$ 3 for fitting. The final result was determined by selecting the model that yielded the lowest $\chi^2$ value.

SED fitting was performed with VOSA using a $\chi^2$-minimization method, with the reduced $\chi^2$ defined as:
\begin{eqnarray}
\chi^2_{\rm r} = \frac{1}{N-n_{\rm p}}\sum_{i=1}^{N}\frac{\left(Y_{i,\rm o}-M_{\rm d}Y_{i,\rm m}\right)^2}{\sigma_{i,\rm o}^2}
\end{eqnarray}
where $Y_{i,\rm o}$ ($Y_{i,\rm m}$) is the observed (model) flux density, $M_{\rm d}$ is the dilution factor, and $\sigma_{i,\rm o}$ is the photometric uncertainty of the $i$-th data point. The degrees of freedom $N-n_{\rm p}$ (where $N$ is the number of valid photometric points and $n_{\rm p}$ is the number of free fitting parameters) are explicitly considered to account for varying $N$ across targets. All photometric errors are propagated throughout the fitting process. For data points with $\sigma_{i,\rm o}=0$, an adjusted uncertainty $\sigma_i=(\delta+0.1)Y_{i,\rm o}$ (where $\delta=\mathrm{Max}(\sigma_i/Y_{i,\rm o})$) is assigned following the VOSA protocol. Data points exhibiting infrared excess are excluded from the fitting, and parameter uncertainties are estimated from model grid steps or Monte Carlo simulations.

Most best-fitting models have solar metallicity and are derived from either BT-Settl or ATMO, which yield similar fitting results. We present ATMO model spectra with fitted SEDs for six objects in Fig.~\ref{fsed}, as these lower-resolution, solar-metallicity models provide optimal visualization and are statistically favoured.

Through SED fitting with VOSA, we obtained atmospheric model spectra that matched the observational data. VOSA successfully identified best-fitting models for 71 of the 768 preliminary candidates. We then used the $T_{\rm eff}$ values inferred from atmospheric models to determine the spectral types of our BD candidates based on established empirical correlations \citep{Kirk21}. We retained only candidates with $T_{\rm eff} < 2300$~K and $3.5 \leq \log g \leq 5.5$ as inferred by their best-fitting models, since this range corresponds to theoretical expectations for BDs \citep{Marl21}. Seventeen objects satisfied our $T_{\rm eff}$ and $\log g$ constraints. We subsequently carried out visual inspections of their images, and three objects were removed owing to very poor F444W band detections.

\begin{table*}
 \centering
    \caption{\textit{JWST} AB magnitudes of three known BDs and three new BD candidates in the UNCOVER field.    }
    \label{tmag}
    \begin{tabular}{lcc ccc ccc}
    \hline\hline
    UNCOVER ID & F090W & F115W & F140M & F150W & F162M & F182M & F200W & F210M  \\ 
    \hline
    UNCOVER-DR3-44489 & --- & 28.12 $\pm$ 0.15 & --- & 27.91 $\pm$ 0.12 & --- & --- & 27.87 $\pm$ 0.13 & ---  \\ 
    UNCOVER-DR3-45661 &  29.30 $\pm$ 0.22 & 27.29 $\pm$ 0.05 & 29.67 $\pm$ 0.75 & 28.08 $\pm$ 0.09 & 27.44 $\pm$ 0.11 & 29.39 $\pm$ 0.35 & 28.75 $\pm$ 0.16 & 28.02 $\pm$ 0.15 \\ 
    UNCOVER-DR3-51651 & --- & 28.03 $\pm$ 0.10 & --- & 28.83 $\pm$ 0.26 & --- & --- & 29.72 $\pm$ 0.44 & ---  \\ 
    UNCOVER-DR3-7408 & 29.80 $\pm$ 0.52 & 27.35 $\pm$ 0.07 & --- & 28.28 $\pm$ 0.17 & --- & --- & 29.07 $\pm$ 0.21 & ---\\ 
    UNCOVER-DR3-8009 & 29.61 $\pm$ 0.29 & 27.80 $\pm$ 0.06 & --- & 28.42 $\pm$ 0.13 & --- & --- & 29.26 $\pm$ 0.16 & --- \\ 
    UNCOVER-DR3-59579 & 28.91 $\pm$ 0.33 & 28.07 $\pm$ 0.14 & --- & 27.67 $\pm$ 0.14 & --- & --- & 28.17 $\pm$ 0.15 & ---  \\ 
    \hline
    \hline
    Name & F277W &  F300M & F335M & F356W & F410M &  F444W & F460M & F480M \\
    \hline
    UNCOVER-DR3-44489 & 28.02 $\pm$ 0.10 &  --- & --- & 27.44 $\pm$ 0.04 &  26.95 $\pm$ 0.05 &  27.19 $\pm$ 0.06 & --- & --- \\ 
    UNCOVER-DR3-45661  &  29.00 $\pm$ 0.15 & 28.77 $\pm$ 0.20 & 29.47 $\pm$ 0.36 &   26.69 $\pm$ 0.03 & 26.80 $\pm$ 0.08 & 26.95 $\pm$ 0.03 & 27.41 $\pm$ 0.16 & 27.22 $\pm$ 0.12 \\ 
    UNCOVER-DR3-51651 &  29.24 $\pm$ 0.23 &  --- & --- & 27.69 $\pm$ 0.05 &  26.02 $\pm$ 0.18 & 25.61 $\pm$ 0.01 & 25.88 $\pm$ 0.25 & --- \\ 
    UNCOVER-DR3-7408  & 28.83 $\pm$ 0.13 & --- & --- & 27.46 $\pm$ 0.04 & 25.54 $\pm$ 0.02 & 25.58 $\pm$ 0.01 & --- & 25.69 $\pm$ 0.05 \\ 
    UNCOVER-DR3-8009  & 29.07 $\pm$ 0.14 & --- & --- & 27.72 $\pm$ 0.04 & 26.17 $\pm$ 0.03 & 26.45 $\pm$ 0.02 & --- & 26.67 $\pm$ 0.11 \\ 
    UNCOVER-DR3-59579  & 28.92 $\pm$ 0.27 & --- & --- & 28.04 $\pm$ 0.14 & --- & 27.90 $\pm$ 0.11 & --- & --- \\ 
    \hline
    \end{tabular}
\end{table*}

\begin{table*}
 \centering
    \caption{VOSA and \textsc{eazy} SED fitting results of UNCOVER BDs to ATMO models and galaxy templates. }
    \label{tprop}
    %\renewcommand
    %\arraystretch{1.2}
    \centering
    \begin{tabular}{llcrlrcc}
    \hline\hline
         UNCOVER ID & Name &   SpT$^a$ &  $T_{\rm eff}$ (K) &  $\log g$   & $\chi^2_{\rm V}$ & $z$ & $\chi^2_{\rm E}$ \\
    \hline
        UNCOVER-DR3-44489 & UNCOVER-BD-1$^b$ & L5.1$^{+1.5}_{-1.3}$ & 1600 $\pm$ 134 & 5.0  & 14.8 & 6.97 & 67.4 \\ 
        UNCOVER-DR3-45661 & UNCOVER-BD-2$^b$ &  T5.8$\pm$0.6 & 1000 $\pm$ 93 & 5.5   &  1.3 & 7.32 & 214.6 \\ 
        UNCOVER-DR3-51651 & UNCOVER-BD-3$^b$ &  T9.6$^{+0.8}_{-0.7}$ & 500 $\pm$ 82 & 5.0 &  3.5 & 1.20 & 111.0 \\
        UNCOVER-DR3-7408 & UNCOVER-BD-4 & T9.1$^{+0.8}_{-0.7}$ & 550 $\pm$ 82 & 5.0 &  6.7 & 1.10 & 216.5 \\ 
        UNCOVER-DR3-8009 & UNCOVER-BD-5 & T7.9$\pm$0.7 & 700 $\pm$ 93 & 5.5 &  5.6 & 7.19 & 133.3 \\ 
        UNCOVER-DR3-59579 & --- &  L5.1$^{+1.8}_{-1.5}$ & 1600 $\pm$ 167 & 5.0   & 0.7 & 6.97 & 16.2 \\ 
    \hline
    \end{tabular}
    \begin{list}{}{}
\item[]$^a$Based on $T_{\rm eff}$ in Fig. \ref{fsed} and spectral type -- $T_{\rm eff}$ correlation \citep[table 13,][]{Kirk21}.
\item[]$^b$Known BDs confirmed by NIRSpec spectroscopy \citep{Bur24}. 
\end{list}
\end{table*}

\begin{table*}
    \centering
    \caption{Best fitting LOWZ models to SEDs of UNCOVER BDs. The log $K_{zz}$ is fixed to 2.0.}
    \label{tlowz}
    \begin{tabular}{lccrr llc ccc}
    \hline\hline
         Name & RA &  DEC &  $T_{\rm eff}$ (K) & [M/H] & log $g$ & C/O & $\chi^2_{\rm L}$ & SpT &  $^a$d (pc) & $^b$Z (pc)\\
        \hline
        UNCOVER-BD-1 & 3.537529188 & $-$30.37016711 & 1400  & $-1.0$ & 5.0 & 0.1 & 4.75 & $^c$sdT1$\pm$1 & 2559$^{+311}_{-277}$ & $-2102^{-255}_{+228}$\\ 
        UNCOVER-BD-2 & 3.546419158 & $-$30.36624575 &1000  & 0.0 & 5.25 & 0.55 & 1.82 & $^c$T6$\pm$0.5 & 1426$^{+187}_{-165}$ & $-1171^{-154}_{+136}$\\ 
        UNCOVER-BD-3 & 3.513893611 & $-$30.35602421 & 500  & 0.0 & 5.0 & 0.55 & 4.38 & $^c$T8.5$\pm$1 & 535$^{+84}_{-73}$ & $-439^{-69}_{+60}$\\ 
        UNCOVER-BD-4 & 3.606273944 & $-$30.44019765 & 600  & 0.0 & 5.25 & 0.55 & 7.32 & $^d$T8$\pm$1 &  527$^{+66}_{-59}$ & $-433^{-54}_{+48}$\\
        UNCOVER-BD-5 & 3.610599496 & $-$30.43831400 & 750  & $-0.3$ & 5.25 & 0.55 & 6.57 & $^d$T7$\pm$1 & 1052$^{+135}_{-119}$ & $-864^{-111}_{+98}$\\
        UNCOVER-DR3-59579 & 3.510205211 & $-$30.34087041 & 1600 & $-0.9$ & 5.0 & 0.55 & 0.65 & $^e$L7$\pm$1 & 3311$^{+607}_{-513}$ & $-2719^{-498}_{+421}$\\
        \hline
    \end{tabular} 
    \begin{list}{}{}
\item[]$^a$Based on correlations between spectral types and F444W-band absolute magnitudes \citep{Bur24}.
\item[]$^b$Galactic height. 
\item[]$^c$Classified by NIRSpec spectra \citep{Bur24}.
\item[]$^d$Derived by F277W $-$ F444W versus spectral type correlation in Fig. \ref{fsptcolor}(b). 
\item[]$^e$Derived by spectral type -- $T_{\rm eff}$ correlation for L subdwarfs \citep{prime3}.
\end{list}
\end{table*}

\iffalse
\begin{table}
    \centering
    \caption{We present surface density of BDs in UNCOVER field with simulations based on the \citep[][B01]{Burr01} and \citep[][B03]{Bara03} evolutionary models.}
    \begin{tabular}{cccc}
    \hline\hline
    SpT & B01 & B03 & UNCOVER \\
    \hline
    L & 0.1041 & 0.1186 & 0.0366\\
    T & 0.1758 & 0.1732 & 0.0914\\
    \hline
    \end{tabular}
    \label{surf-den}
\end{table}
\fi

\subsection{NIRSpec spectra validation}
\label{ssspec}

Among our 14 BD candidates, UNCOVER-DR3-45661 and UNCOVER-DR3-51651 have been spectroscopically confirmed as BDs through NIRSpec observations, designated as UNCOVER-BD-2 (T6) and UNCOVER-BD-3 (T8–T9), respectively \citep{Bur24}. The best-fitting stellar models to the SEDs of UNCOVER-BD-2 (Fig. \ref{fsed}, middle-left panel) and UNCOVER-BD-3 (Fig. \ref{fsed}, top-left panel) provide significantly better fits than their best-fitting galaxy templates (see Table~\ref{tprop}).

UNCOVER-BD-2 exhibits a well-defined F410M band peak characteristic of cool BDs, enabled by high-quality multi-band detections spanning 3–5~$\mu$m. The excellent agreement between the observed SED and theoretical models across this broad wavelength range demonstrates the power of \textit{JWST}'s multi-band photometry for identifying and characterizing cool substellar objects. UNCOVER-BD-3 similarly shows a robust SED model fit with reliable detections and a sufficiently cool $T_{\rm eff}$ to reveal both the V-shaped feature centred at F277W and the characteristic F410M band peak.

UNCOVER-BD-1 \citep[sdT1;][]{Bur24} was not recovered in our initial search, as its F277W $-$ F444W = 0.97 colour falls marginally below our selection threshold (Equation \ref{eq1}). Fig. \ref{fcolor} illustrates that models with [M/H] = $-0.5$ exhibit slightly bluer F277W $-$ F444W colours compared to solar-metallicity models. For comparison, we fitted UNCOVER-BD-1's SED using VOSA following the same procedure applied to our other candidates. Fig. \ref{fsed} (bottom-left panel) shows UNCOVER-BD-1's SED with its best-fitting solar-metallicity model ($T_{\rm eff}$ = 1600K, $\log g$ = 5). This SED fitting result differs significantly from the best-fitting LOWZ model to its NIRSpec spectrum ($T_{\rm eff}$ = 1300K, $\log g$ = 5.0, [M/H] = $-1.0$; \citealt{Bur24}). This discrepancy arises because UNCOVER-BD-1 is a metal-poor BD, whereas ATMO models are limited to solar metallicity. Consequently, ATMO models favour higher $T_{\rm eff}$ values to compensate for metal-poor spectral features. This finding suggests that our SED fitting may overestimate $T_{\rm eff}$ for metal-poor BDs, potentially causing early-type T subdwarfs to be missed or misclassified as warmer L dwarfs in colour- or SED-based searches. SEDs may not contain sufficient information to disentangle early-T subdwarfs from late-L dwarfs through model fitting.

To validate our SED-based selection, we searched for spectra of the remaining twelve BD candidates in the UNCOVER catalogue through the Mikulski Archive for Space Telescopes \citep[MAST;][]{mast}. Four candidates have been observed with NIRSpec in prism mode: UNCOVER-DR3-9334, UNCOVER-DR3-10065, UNCOVER-DR3-12259, and UNCOVER-DR3-41096 (Table~\ref{tgala}). We extracted their NIRSpec spectra using the NIRSpec analysis tool \textsc{msaexp} \citep{bram22}. 

Fig. \ref{fgspec} presents the NIRSpec spectra of these four objects. Redshifted absorption features arising from the Lyman-$\alpha$ forest and Balmer break in their spectra reveal that all four are high-redshift galaxies \citep[e.g. fig. 17,][]{wold25}. Ly$\alpha$$\lambda1216$, Ne\textsc{iii}$\lambda3869$, H$\gamma$, H$\beta$, and [O\textsc{iii}]$\lambda5007$ appear in UNCOVER-DR3-9334. Ly$\alpha$, H$\gamma$, H$\beta$, H$\alpha$, and [O~\textsc{iii}]$\lambda5007$ appear in both UNCOVER-DR3-10065 and UNCOVER-DR3-41096. UNCOVER-DR3-41096 also exhibits a reddened component beyond 3$\mu$m. UNCOVER-DR3-12259 has a lower redshift and shows only the Balmer break. UNCOVER-DR3-10065 and UNCOVER-DR3-41096 were reported as unconfirmed broad-line active galactic nuclei (AGN) under the designations UNCOVER-DR1-571 and UNCOVER-DR1-20080, with redshifts of 6.74 and 7.04, respectively \citep{gree24}.

We measured the redshifts of these four galaxies using the identified spectral features. UNCOVER-DR3-9334 and UNCOVER-DR3-12259 have redshifts of 7.19 and 2.62, respectively, according to our measurements. We also confirmed the redshifts of UNCOVER-DR3-10065 and UNCOVER-DR3-41096 as measured by \citet{gree24}.

\subsection{Comparison of stellar and galactic SED fitting}
As demonstrated in Section~\ref{ssspec}, our candidate list is contaminated by high-redshift galaxies exhibiting colours similar to those of BDs. To identify additional potential contaminants among the remaining eight candidates lacking NIRSpec spectra, we compared the $\chi^2_{\rm V}$ values from our VOSA stellar model SED fitting (Section~\ref{ssvosa}) with the $\chi^2_{\rm E}$ values from \textsc{eazy} galactic template SED fitting in the SUPER catalogue of UNCOVER DR3 \citep{Sues24}.

The $\chi^2_{\rm V}$ and $\chi^2_{\rm E}$ values of the best-fitting stellar models and galactic templates are presented in Table~\ref{tgala}. Fig.~\ref{fgal} shows the best-fitting stellar models (ATMO) and galaxy templates for five objects. We employed \textsc{eazy} to fit the SEDs of these objects using the fixed redshifts from UNCOVER DR3 to obtain the best-fitting galaxy templates.

Our analysis revealed that one of these eight candidates (UNCOVER-DR3-68341) was significantly better fitted by galaxy templates than by stellar models ($\chi^2_{\rm E} < 0.5\chi^2_{\rm V}$), indicating a higher likelihood of being a galaxy rather than a BD. We therefore excluded UNCOVER-DR3-68341 from our BD candidate list. The best-fitting galaxy template yields a $\chi^2_{\rm E}$ value of 0.3, substantially lower than the $\chi^2_{\rm V}$ value of 9.3 for the best-fitting stellar model. This result strongly suggests that UNCOVER-DR3-68341 is a galaxy at a redshift of $z \sim 7.69$.

UNCOVER-DR3-72947 and UNCOVER-DR3-3847 (Fig.~\ref{fgal}, middle panels) were marginally better fitted by galactic templates ($\chi^2_{\rm E}$ = 4.1, 4.0) than by stellar models ($\chi^2_{\rm V}$ = 5.3, 6.1). However, these differences are not statistically significant enough to draw definitive conclusions for these two objects.

While UNCOVER-DR3-13700 and UNCOVER-DR3-15552 (Fig.~\ref{fgal}, bottom panels) showed significantly better fits to stellar models ($\chi^2_{\rm V}$ = 1.3, 1.4) than to galaxy templates ($\chi^2_{\rm E}$ = 19.4, 70.1), both sources suffer from poor detection quality in multiple NIRCam images, as illustrated in the bottom panels of Fig.~\ref{fgal}. Given these unreliable photometric measurements, we removed them from our BD candidate list to ensure the robustness of our final sample.

The remaining three candidates (UNCOVER-DR3-7408, UNCOVER-DR3-8009, and UNCOVER-DR3-59579) were significantly better fitted by stellar models than by galaxy templates, suggesting a higher likelihood of being BDs rather than galaxies. These objects are further discussed in Section~\ref{results}.

\begin{figure*}
    \includegraphics[width=\textwidth]{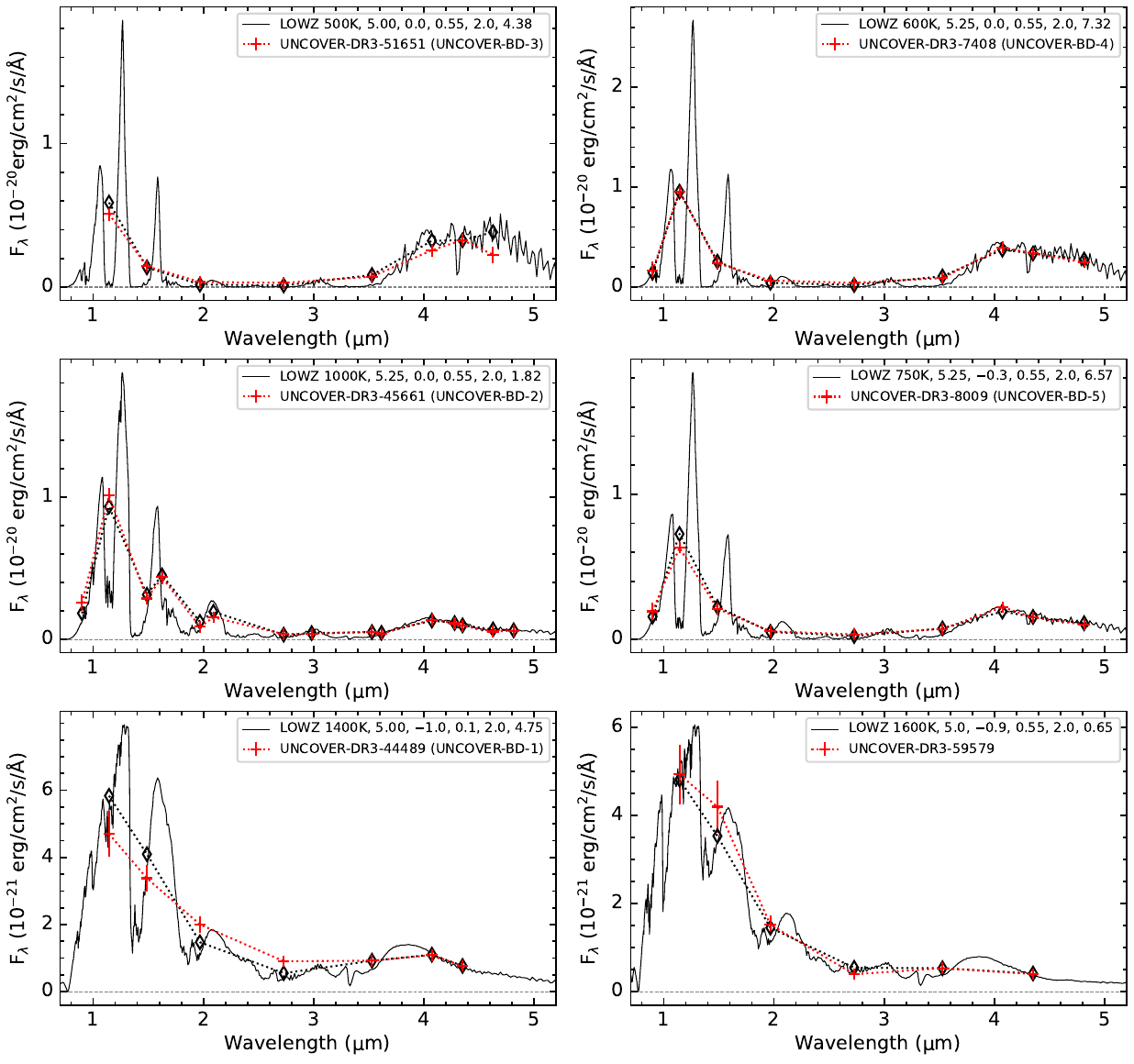}
    \caption{Best-fitting LOWZ models (black lines and diamonds) to the SEDs (red crosses) of three known BDs (left panels) and three BD candidates (right panels). The model parameters $T_{\rm eff}$, $\log g$, [M/H], C/O, and $\log_{10}K_{zz}$ (fixed at 2.0), along with the $\chi^2_{\rm L}$ values of the fits, are indicated.
    }
    \label{flowz}
\end{figure*}

\section{Brown dwarf candidates}
\label{results}
After applying our complete selection criteria, we identified three new BD candidates. \textit{JWST} photometry for these three candidates, along with the three previously known UNCOVER BDs, is presented in Table~\ref{tmag}. Fig.~\ref{fsed} (right panels) displays the SED fits and corresponding \textit{JWST} images for the three new BD candidates, while Table~\ref{tprop} summarizes the atmospheric parameters derived from their best-fitting ATMO models. We determined spectral types for these candidates by applying the $T_{\rm eff}$–spectral type relation from \citet{Kirk21} to the $T_{\rm eff}$ values obtained from our SED fitting. This analysis yielded two candidates with $T_{\rm eff}$ consistent with T dwarfs and one consistent with L dwarfs.

\begin{figure*}
    \centering    
    \includegraphics[width=\linewidth]{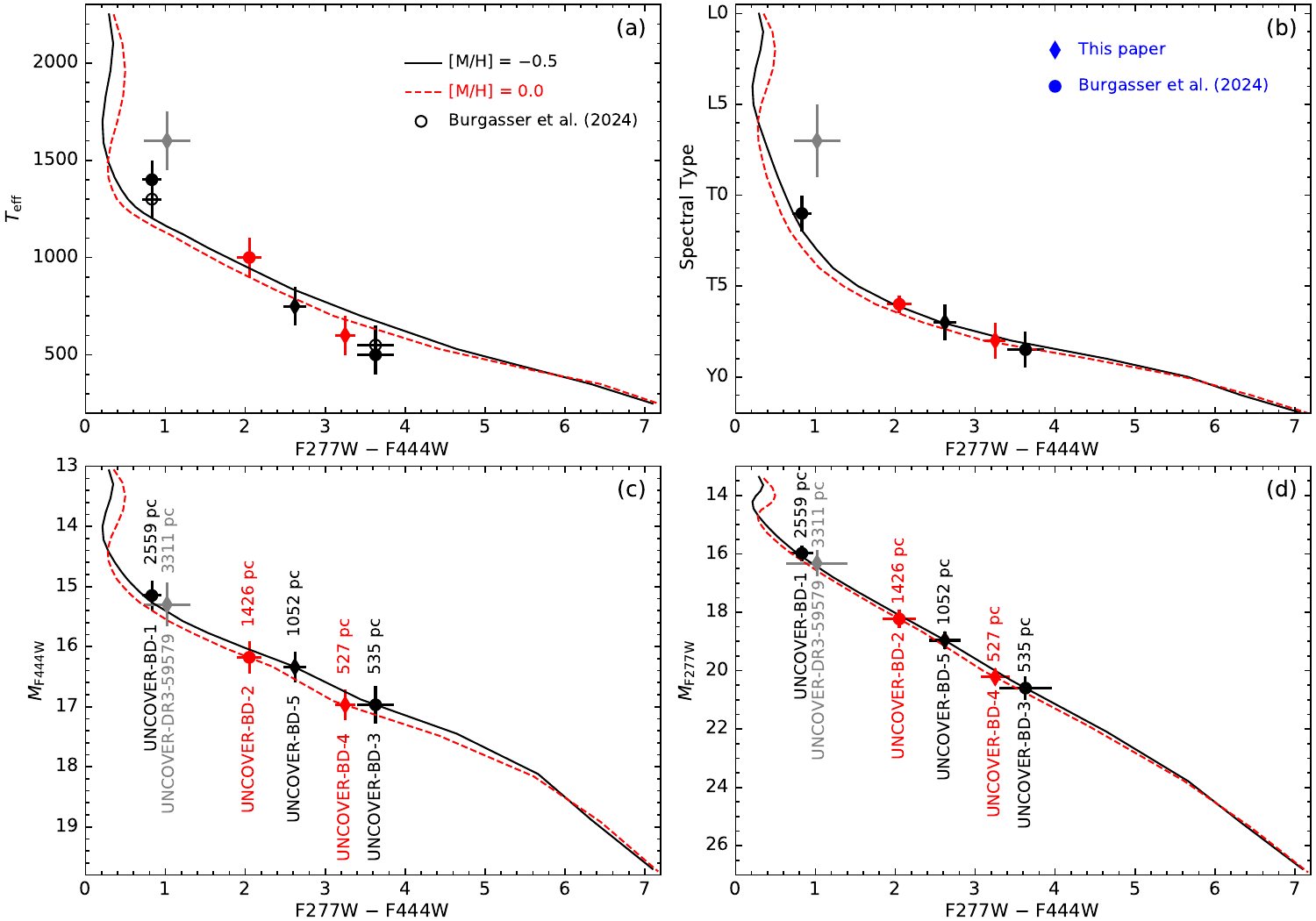}
    \caption{Model-predicted correlations between $T_{\rm eff}$, spectral type, $M_{\rm F444W}$, $M_{\rm F277W}$, and F277W$-$F444W colour \citep[table~2 of][]{Bur24}. Black solid lines indicate [M/H] $= -0.5$, while red dashed lines indicate [M/H] $= 0$. BDs from Table~\ref{tmag} are overplotted as circles \citep[from][]{Bur24} and diamonds (this work). BDs with solar and subsolar metallicities are plotted in red and black, respectively. Open circles indicate alternative $T_{\rm eff}$ values derived from spectral fitting \citep{Bur24}.
    }
    \label{fsptcolor}
\end{figure*}

Fig.~\ref{fcolor} shows the locations of our three new BD candidates and the three known UNCOVER BDs in colour–colour space, demonstrating how they populate the region defined by our selection criteria. Fig.~\ref{fsptcolor} shows the spectral type versus F277W $-$ F444W colour correlation for the three known BDs and the three new BD candidates compared to predictions \citep[table 2 of][]{Bur24} based on atmospheric models with [M/H] = 0 and [M/H] = $-$0.5 \citep{Marl21}. The successful recovery of UNCOVER-BD-2 and UNCOVER-BD-3 validates our colour selection methodology, while the identification of three additional candidates with photometric properties consistent with substellar objects suggests that deeper \textit{JWST} surveys will continue to reveal the faint, distant BD population.

\subsection{Two robust T dwarf candidates}
UNCOVER-DR3-7408 (hereafter UNCOVER-BD-4) represents the coolest BD candidate in our sample (Fig.~\ref{fsed}, top-right panel). The object is detected across nine NIRCam bands from F090W to F480M (Table~\ref{tmag}) and exhibits the highest mid-infrared flux among our candidates in the F277W–F480M range. The best-fitting ATMO 2020 model ($\chi^2_{\rm V}$ = 6.7) yields $T_{\rm eff}$ = 550 K, corresponding to a spectral type of T9.1$^{+0.8}_{-0.7}$ for field dwarfs. This stellar model significantly outperforms the \textsc{eazy} galaxy fit ($z$ = 1.10, $\chi^2_{\rm E}$ = 216.5), strongly supporting a BD interpretation. The object's prominent mid-infrared excess closely resembles the spectral characteristics of confirmed late-T dwarfs such as UNCOVER-BD-3 \citep{burg02}, leading us to designate it as UNCOVER-BD-4. 

UNCOVER-DR3-8009 (hereafter UNCOVER-BD-5) also exhibits robust detections across nine NIRCam bands from F090W to F480M (Table~\ref{tmag}). The best-fitting ATMO 2020 model ($\chi^2_{\rm V}$ = 5.6) indicates $T_{\rm eff}$ = 700 K, providing a substantially better fit than the \textsc{eazy} galaxy template ($z$ = 7.19, $\chi^2_{\rm E}$ = 133.3). The object's SED, particularly its enhanced mid-infrared flux, displays the characteristic profile of a T dwarf (Fig.~\ref{fsed}, middle-right panel). Based on these compelling photometric properties, we designate this object as UNCOVER-BD-5. The $T_{\rm eff}$ (700 K) of UNCOVER-BD-5 corresponds to a spectral type of T7.9 $\pm$ 0.7 for field dwarfs. We note that the bright source west of UNCOVER-BD-5 is a background star at a photometric distance of several tens of kiloparsecs.

\begin{figure}
    \centering    
    \includegraphics[width=0.9\linewidth]{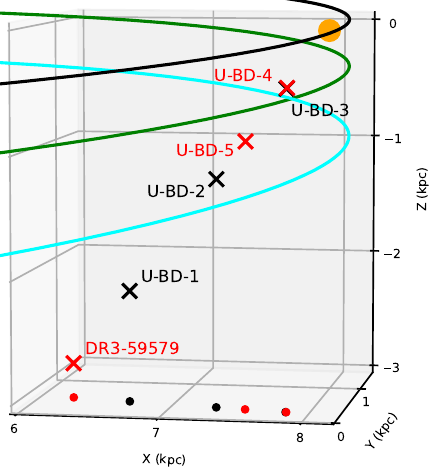}
    \caption{
    Three-dimensional distribution of BDs in Galactic coordinates, showing three known BDs (black crosses) and three new BD candidates (red crosses) identified in the UNCOVER field (see Table~\ref{tprop}). Galactic heights (3.7, 1.9, 0.7~kpc) of known BDs are based on their spectroscopic distances \citep{Bur24}. Galactic heights of new BD candidates are based on their photometric distances (Table~\ref{tprop}). The viewing angle is set at azimuth $= 280\degr$ and elevation $= 6\degr$. Black and red points indicate the projections of these objects on to the Galactic plane ($X$--$Y$). The Galactic Centre is located at the origin (0, 0, 0), while the Sun's position is marked by an orange filled circle at $(8.178, 0, 0.021)$~kpc \citep{grav19}. The inclined black ring with radius 8.178~kpc traces the Solar circle in the Galactic plane. The green and cyan rings mark the lower boundaries of the Galactic thin disc at $Z = -0.4$~kpc and thick disc at $Z = -1.0$~kpc, respectively.}
    \label{fgala}
\end{figure}

\subsection{A possible L subdwarf}
UNCOVER-DR3-59579 shows a significantly better fit to stellar models than to galaxy templates (Fig.~\ref{fsed}, bottom-right panel). Its SED displays a V-shaped profile similar to that of UNCOVER-BD-1, with a flux minimum at F277W. Although the absence of F410M photometry prevents full characterization of the 4~$\mu$m peak, the object exhibits clear detections across most NIRCam bands, supporting its classification as a BD candidate. Note that the F090M band flux is not included in the SED fitting because it appears undetected in the F090W band image.

UNCOVER-DR3-59579 shares notable similarities with UNCOVER-BD-1, the T1 subdwarf with $T_{\rm eff}$ = 1300 K and [M/H] = $-1$ beyond the thick disc \citep{Bur24}. The best-fitting solar-metallicity ATMO model yields $T_{\rm eff}$ = 1600 K for UNCOVER-DR3-59579, corresponding to a spectral type of L5.1$^{+1.5}_{-1.3}$ for field L dwarfs. However, its Galactic height ($Z = -2719^{-498}_{+421}$ pc; Section~\ref{ssdis}) beyond the thick disc and spectral similarities to UNCOVER-BD-1 suggest it may possess subsolar metallicity. Consequently, its true $T_{\rm eff}$ could be cooler than indicated by solar-metallicity models, potentially placing it in the late-L or early-T subdwarf regime.

\subsection{SED fitting to low metallicity models}
\label{slowz}
The newly identified distant UNCOVER BDs may possess subsolar metallicities, as both UNCOVER-BD-1 and UNCOVER-BD-3 are confirmed to be metal-poor. We fitted the SEDs of the UNCOVER BDs listed in Table~\ref{tmag} to LOWZ models, which incorporate subsolar metallicity atmospheres \citep{meis21}. The LOWZ models span a $T_{\rm eff}$ range from 500 to 1600~K. Since the information that can be derived from broad-band SEDs is limited due to their inherently low spectral resolution, we constrained the ranges of several insensitive parameters. We adopted a fixed log~$K_{zz}$ = 2.0, which has a relatively minor impact on spectral profile variations. We restricted our analysis to models with C/O ratios of 0.1 and 0.55, as metal-poor BDs tend to exhibit lower C/O values \citep[e.g. Wolf 1130C;][]{burg25}. We limited the metallicity range to $-2.0 \le$ [M/H] $\le 0.0$, interpolated to steps of 0.1~dex. We employed only models with fixed log~$g$ values of 5.0, 5.25, and 5.5~dex. As distant T~dwarfs, these UNCOVER BDs are very likely more massive than 0.03~M$_{\odot}$ and older than 1~Gyr \citep[e.g. fig.~5;][]{prime6}, implying log~$g \geq 5.0$~dex according to the Sonora Bobcat models \citep{Marl21}.

To fit the observed SEDs of these UNCOVER BDs, we first convolved the LOWZ model spectra through the relevant filter bandpasses to obtain the corresponding synthetic photometry. We then fitted these model SEDs to the observed SEDs of the UNCOVER BDs. The best-fitting model for each object was selected based on the minimum $\chi^2_{\rm L}$ across all fits. Fig.~\ref{flowz} presents the best-fitting LOWZ models to the SEDs of these UNCOVER BDs. Three spectroscopically confirmed UNCOVER BDs are included in our fitting sample for method validation and comparison. The fitting results are presented in Table~\ref{tlowz}.

The SED fitting result ($T_{\rm eff}$ = 1000 K, log $g$ = 5.25, [M/H] = 0.0) by LOWZ models for UNCOVER-BD-2 is consistent to that of spectral fitting \citep{Bur24} and VOSA fitting result by ATMO models (Fig. \ref{fsed}). This is partially due to the wide wavelength coverage of \textit{JWST} filters from F090W to F480M. Mid to late-T dwarfs also have more distinguishable features than earlier type BDs. 

The subsolar metallicity of UNCOVER-BD-3 suggested by spectral fitting \citep{Bur24} is not recovered in our SED fitting to LOWZ models. The best-fitting LOWZ model ($T_{\rm eff}$ = 500~K, log~$g$ = 5.0, [M/H] = 0.0) also favours a slightly cooler temperature.

The best-fitting LOWZ model ($T_{\rm eff}$ = 600~K, log~$g$ = 5.25, [M/H] = 0.0) suggests solar metallicity for UNCOVER-BD-4, which is slightly warmer than UNCOVER-BD-3. The fitting result for UNCOVER-BD-4 should be more reliable than that for UNCOVER-BD-3 owing to its brighter magnitudes and the additional F090W band measurement. The derived $T_{\rm eff}$ of 600~K corresponds to a spectral type of T8.7 according to the spectral type--$T_{\rm eff}$ relation \citep{Kirk21}. This $T_{\rm eff}$ is slightly lower than that indicated by the F277W$-$F444W versus $T_{\rm eff}$ correlation (Fig.~\ref{fsptcolor}(a)). The same discrepancy is observed for UNCOVER-BD-3. The F277W$-$F444W versus spectral type correlation in Fig.~\ref{fsptcolor}(b) suggests that UNCOVER-BD-4 has a spectral type of $\sim$T8, as it exhibits a slightly bluer F277W$-$F444W colour than the T8--T9 type UNCOVER-BD-3. Since F277W$-$F444W colour is a more direct observable than $T_{\rm eff}$, we adopt this spectral type estimate.

UNCOVER-BD-5 also benefits from full F090W to F480M band coverage. The best-fitting LOWZ model ($T_{\rm eff}$ = 750~K, log~$g$ = 5.25, [M/H] = $-0.3$) indicates subsolar metallicity for UNCOVER-BD-5. This metallicity is consistent with thick disc membership, which is further supported by its Galactic height ($Z = -864^{+98}_{-111}$~pc; see Section~\ref{ssdis}). A $T_{\rm eff}$ of 750~K corresponds to a spectral type of T7.5 for field BDs. The F277W$-$F444W versus spectral type correlation in Fig.~\ref{fsptcolor}(b) suggests a T7 classification for UNCOVER-BD-5. Considering the subsolar metallicity inferred from its best-fitting model, we classify UNCOVER-BD-5 as a T7 subdwarf.

The SED fitting to LOWZ models for UNCOVER-BD-1 yields $T_{\rm eff}$ = 1400~K, log~$g$ = 5.25, and [M/H] = 0.0, representing a higher temperature and identical metallicity compared to the spectral fitting result \citep{Bur24}. 
The best-fitting LOWZ model for UNCOVER-DR3-59579 ($T_{\rm eff}$ = 1600~K, log~$g$ = 5.0, [M/H] = $-0.9$) suggests subsolar metallicity, whilst yielding the same $T_{\rm eff}$ as derived from the solar-metallicity ATMO model fitting. This best-fitting model corresponds to a spectral type of sdL7 according to the spectral type--$T_{\rm eff}$ relation for L~subdwarfs \citep[e.g. fig.~4;][]{prime3}. We retain UNCOVER-DR3-59579 as a BD candidate, as the characteristic 4~$\mu$m flux peak cannot be verified due to the absence of F410M band measurements.

\section{Galactic distribution}
\label{dicuss}
\subsection{Photometric distance}
\label{ssdis}
Figure~\ref{fsptcolor}(c) demonstrates that the F277W$-$F444W colour is well correlated with the F444W-band absolute magnitude ($M_{\rm F444W}$). We therefore used the F277W$-$F444W colours of these UNCOVER BDs to derive their $M_{\rm F444W}$ values, and subsequently estimated photometric distances by combining their observed F444W-band magnitudes with the $M_{\rm F444W}$ corresponding to their F277W$-$F444W colours. These photometric distances are indicated in Fig.~\ref{fsptcolor}(c) and listed in Table~\ref{tlowz}.

The distance distribution of our candidates spans approximately 0.5–3.3~kpc and exhibits an inverse correlation with $T_{\rm eff}$. This trend reflects the survey's sensitivity limits, which constrain the maximum detectable distance for BDs and decrease for cooler objects. Fig.~\ref{fgala} illustrates the Galactic distribution of these candidates relative to the Galactic Centre, the Sun, and the Milky Way's thin and thick discs. The two robust T dwarf candidates, UNCOVER-BD-4 and UNCOVER-BD-5, have Galactic heights of $Z = -433^{-54}_{+48}$ and $-864^{-111}_{+98}$~pc, respectively, placing them near the outer boundaries of the thin and thick discs. Notably, the L subdwarf candidate UNCOVER-DR3-59579 lies far beyond the thick disc boundary. The metallicities of these three new BD candidates, as inferred from the best-fitting LOWZ models, are consistent with their membership in the Galactic thin disc, thick disc, and halo, respectively.

\begin{figure}
    \centering    
    \includegraphics[width=\linewidth]{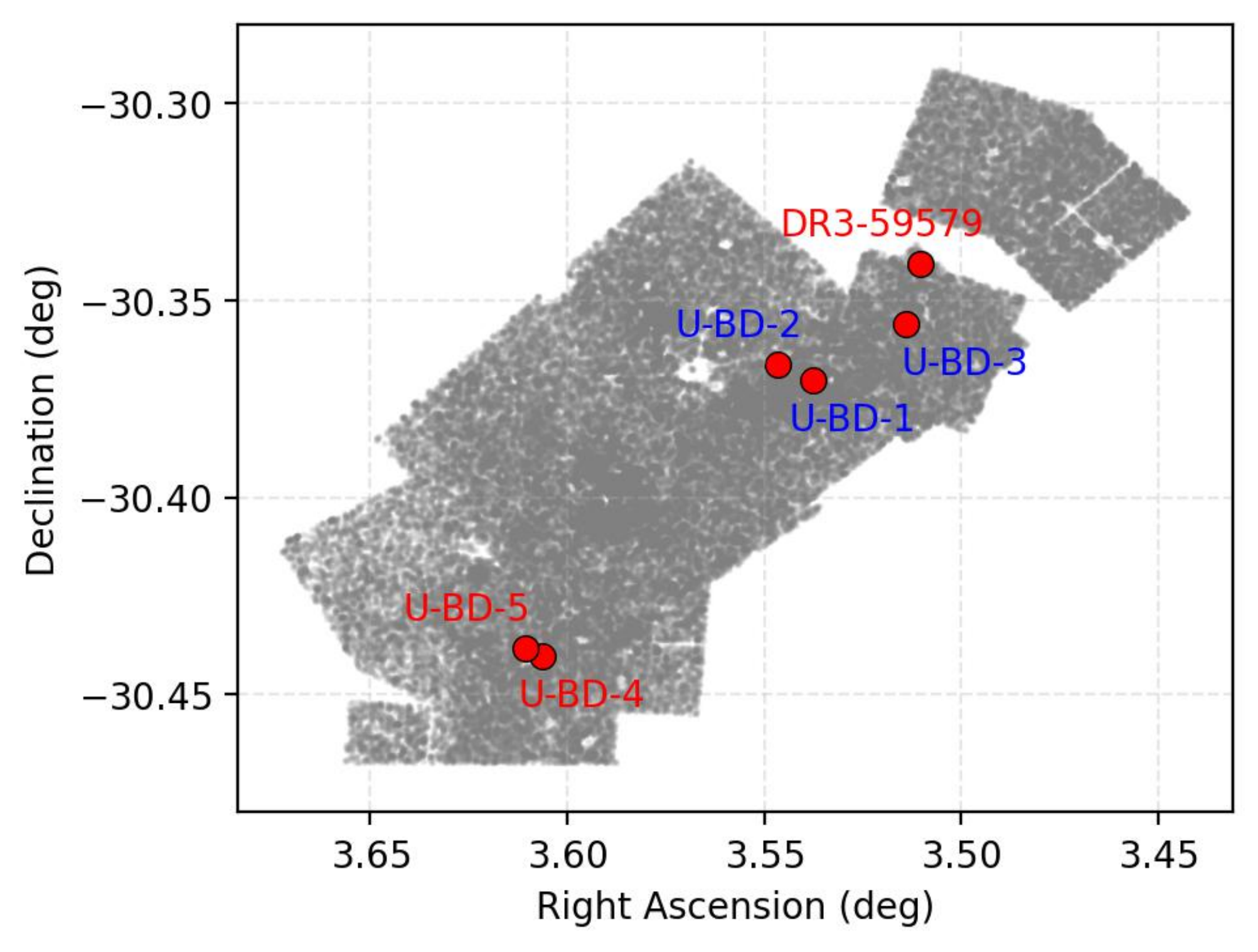}
    \caption{Common footprint (grey dots) with F277W, F444W, and F115W/F150W band photometry in the `SUPER' catalogue of UNCOVER and MegaScience DR3. Known BDs and new candidates in Table~\ref{tprop} are highlighted with red dots.}
    \label{funcover}
\end{figure}

\subsection{Survey area and depth}
\label{sssden}
To estimate the surface and space density of BDs in UNCOVER, we computed the footprint of the common area covered by the F277W, F444W, and F115W/F150W band mosaic images in DR3 of the UNCOVER and MegaScience surveys. Coverage in these bands is necessary for our initial colour selection (Fig.~\ref{fcolor}). We quantified the footprint of the DR3 mosaic images via FITS header metadata and pixel data properties \citep{Bezan24,Sues24}. The per-pixel solid angular area was calculated from the determinant of the WCS CD matrix elements in the FITS primary header, which describe the linear transformation from pixel coordinates to celestial coordinates. As the UNCOVER observations are arranged in a mosaic containing some unobserved/blank pixels, we masked the imaging array to retain only valid observed pixels. We multiplied the number of valid pixels by the single-pixel angular area, obtaining a total effective sky coverage of 53.4~arcmin$^2$ (Fig. \ref{funcover}).

Reliable identification of photometric candidates of T dwarfs in UNCOVER requires robust detections across multiple NIRCam bands, particularly in F277W where T dwarfs are faintest. Consequently, we define the survey depth using the F277W magnitude of the faintest object in our sample (UNCOVER-BD-3 with F277W = 29.24~AB~mag) rather than the nominal UNCOVER limiting magnitudes. Fig.~\ref{fsptcolor}(d) shows the F277W$-$F444W versus $M_{\rm F277W}$ correlation, with UNCOVER BDs overplotted using F277W-band magnitudes adjusted for their respective distances.

\subsection{Surface and space number densities}
\label{ssnden}
With the five T dwarfs identified in the UNCOVER field—two from this work (UNCOVER-BD-4, UNCOVER-BD-5) and three from previous studies (UNCOVER-BD-1, UNCOVER-BD-2, UNCOVER-BD-3), we determine a surface density of 0.094~arcmin$^{-2}$ over a covered area of 53.4~arcmin$^2$, to a limiting depth of F277W $= 29.24$~mag (or F115W $= 28.03$~mag, based on UNCOVER-BD-3).
This result can be compared with predictions by \citet{agan22b}, who employed the evolutionary models of \citet{Burr01} and \citet{Bara03} to estimate surface densities of 0.18 and 0.17~arcmin$^{-2}$ for field T dwarfs out to 400~pc in the \textit{JWST}/PASSAGE survey fields \citep{runn25}, assuming a limiting depth of F115W $= 27$~mag (AB). However, direct comparison requires careful consideration of several factors, including Galactic latitude, Galactic height distribution, and survey depth.

Our measured surface density is approximately a factor of two lower than the \citet{agan22b} predictions, despite our survey being one magnitude deeper in the F115W band. This apparent discrepancy can be attributed to differences in the sampled Galactic heights: whilst the \citet{agan22b} predictions are limited to distances within 400~pc of the Galactic plane, all T dwarfs in our sample have estimated Galactic heights beyond 400~pc, where the stellar density is expected to decline. Additional systematic uncertainties in the \citet{agan22b} predictions may arise from the adopted Galactic density parametrization, the evolutionary models employed, and the assumed absolute magnitude--spectral type relations.
Despite these differences, our measurement remains consistent with the \citet{agan22b} predictions at the order-of-magnitude level, particularly when accounting for the substantial uncertainties arising from small number statistics.

\begin{table}
    \centering
    \caption{Space density of T5+ dwarfs based on two new robust BD candidates and two known BDs in the UNCOVER field.}
    \label{den}
    \begin{tabular}{lccccc}
    \hline\hline
        $T_{\rm eff}$  & $^ad_{\rm max}$ & $Z_{\rm max}$ & Count & $^b$Density  & $^c$Density-K\\
        (K)  & (kpc) & (kpc) & & ($10^{-3}$ pc$^{-3}$) &   ($10^{-3}$ pc$^{-3}$) \\
        \hline
        (900, 1050] & 2.6  & 2.2 & 1 & 0.04 & 1.72  \\ 
        (750, 900]  & 2.0  & 1.6 & 0 & --- & 1.99\\ 
        (600, 750]  & 1.3  & 1.1 & 1 & 0.30 & 2.80 \\ 
        (450, 600]  & 0.7  & 0.6 & 2 & 3.76 & 4.24 \\  
        \hline
        SpT & ~ & ~ & ~ & ~ \\ 
        \hline
        T5-5.9 & 2.6 & 2.2 & 0 & --- & 1.18 \\ 
        T6-6.9  & 2.0 & 1.6 & 1 & 0.09 & 1.45 \\ 
        T7-7.9  & 1.3  & 1.1 & 1 & 0.30 & 1.52 \\ 
        T8-8.9  & 0.7 & 0.6 & 2 & 3.76 & 3.54 \\
        T9-9.9  & 0.3 & 0.2 & 0 & --- & 1.86 \\   
        \hline
    \end{tabular} \\
    \begin{list}{}{}
    \item[]Notes.  $^a$The survey depth is defined by the faintest object in the sample (F277W = 29.24 for UNCOVER-BD-3).
    $^b$Results from this paper.  $^c$Density based on BD sample within 20 pc of the Sun \citep[table 15, ][]{Kirk21}.
    \end{list}
\end{table}

Table~\ref{den} compares the space density of T5-T9 dwarfs within the UNCOVER detection cone -- including UNCOVER-BD-2 and UNCOVER-BD-3 from \citet{Bur24} -- with the local population within 20~pc \citep{Kirk21}, as a function of $T_{\rm eff}$ and spectral type. 
The number density of T5–T7 dwarfs in the UNCOVER cone extending to 2.3$\sim$2.0~kpc is substantially lower than in the 20~pc sample, with only UNCOVER-BD-2 representing this spectral range. Similarly, the T7 dwarf density out to $\sim$1.5~kpc falls significantly below solar neighbourhood values, represented solely by UNCOVER-BD-5. 
In contrast, the T8$\-$T8.9 dwarf density appears comparable to local values. This suggests that the number density of T8-T8.9 dwarfs remains relatively constant throughout the thick disc.
This trend may partially reflect the extended cooling timescales of thick disc BDs, most of which have evolved to late-T spectral types \citep[e.g., fig.~5 in][]{prime6}.

\section{Summary and conclusions}
\label{conclusion}
We conducted a systematic search for BD candidates in the \textit{JWST} UNCOVER and MegaScience surveys. Our methodology comprised initial colour-based selection, followed by SED fitting to stellar models using VOSA and to high-redshift galaxy templates using \textsc{eazy}. Through archival searches on MAST, we identified four high-redshift galaxies via their NIRSpec spectra. Notably, the SEDs of these four galaxies showed better fits to stellar models than to galaxy templates, demonstrating the challenge of distinguishing BDs with $T_{\rm eff} \gtrsim 900$~K (spectral types earlier than T5) from high-redshift galaxies using NIRCam photometry alone.

Our analysis identified a robust T8 dwarf candidate (UNCOVER-BD-4), a robust T7 subdwarf candidate (UNCOVER-BD-5), and a possible L subdwarf candidate (UNCOVER-DR3-59579) in the UNCOVER DR3 catalogues. The two T dwarfs reside near the boundary between the Galactic thin and thick discs. UNCOVER-DR3-59579, with a Galactic height of 2.7~kpc, lies well within the Galactic halo. UNCOVER-DR3-59579 exhibits SED features similar to UNCOVER-BD-1. Best-fitting LOWZ model suggesting it may be an late-L subdwarf.

In the UNCOVER field, we determine a T dwarf surface density of 0.094~arcmin$^{-2}$ to a limiting depth of F115W = 28.03~mag, lower than previous predictions but consistent at the order-of-magnitude level.
We also calculated BD space number densities within the UNCOVER detection cone. The number density of T5–T6.9 dwarfs extending to 2.2$\sim$1.6~kpc falls substantially below solar neighbourhood values. The T7 dwarf density out to 1.1~kpc, while slightly higher, remains significantly lower than local measurement. Conversely, the number density of T8–T8.9 dwarfs within the thick disc appears comparable to solar neighbourhood values.

This study demonstrates the capability of \textit{JWST} deep surveys to identify distant, cool substellar objects. Spectroscopic confirmation and characterization of these candidates will refine distance constraints and provide crucial insights into the Galactic distribution and properties of BDs, advancing our understanding of substellar formation and evolutionary processes.

\section*{Acknowledgements}
This work is based on observations made with the NASA/ESA/CSA \textit{JWST}. The data were obtained from the Mikulski Archive for Space Telescopes at the Space Telescope Science Institute, which is operated by the Association of Universities for Research in Astronomy, Inc., under NASA contract NAS 5-03127 for JWST. 
These observations are associated with programs JWST-GO-2561, JWST-GO-2883, JWST-GO-3516, JWST-GO-3538, JWST-GO-4111, JWST-ERS-1324, and JWST-DD-2767.
This publication makes use of VOSA, developed under the Spanish Virtual Observatory project supported from the Spanish MICINN through grant AyA2008-02156. 
MCGO acknowledges financial support from the Agencia Estatal de Investigación (AEI/10.13039/501100011033) of the Ministerio de Ciencia e Innovación and the ERDF “A way of making Europe” through project PID2022-137241NB-C42. The authors thank the referee for the useful and constructive comments.

%%%%%%%%%%%%%%%%%%%%%%%%%%%%%%%%%%%%%%%%%%%%%%%%%%
\section*{Data Availability}

The data underlying this article are available in the article and in its online supplementary material.

%Check Sample Data Availability Statements at \url{https://academic.oup.com/pages/open-research/research-data#Data%20Availability%20Statements}.

%The inclusion of a Data Availability Statement is a requirement for articles published in MNRAS. Data Availability Statements provide a standardised format for readers to understand the availability of data underlying the research results described in the article. The statement may refer to original data generated in the course of the study or to third-party data analysed in the article. The statement should describe and provide means of access, where possible, by linking to the data or providing the required accession numbers for the relevant databases or DOIs.



\bibliographystyle{mnras}
\bibliography{example} % if your bibtex file is called example.bib

%\begin{thebibliography}{99}
%\bibitem[\protect \citeauthoryear{Author}{2012}]{Author2012}
%\bibitem[\protect \citeauthoryear{Others}{2013}]{Others2013}
%\end{thebibliography}

%%%%%%%%%%%%%%%%%%%%%%%%%%%%%%%%%%%%%%%%%%%%%%%%%%

%%%%%%%%%%%%%%%%% APPENDICES %%%%%%%%%%%%%%%%%%%%%

\appendix
\section{High-redshift galaxies and ambiguous objects}
\begin{figure*}
    \includegraphics[width=0.495\textwidth]{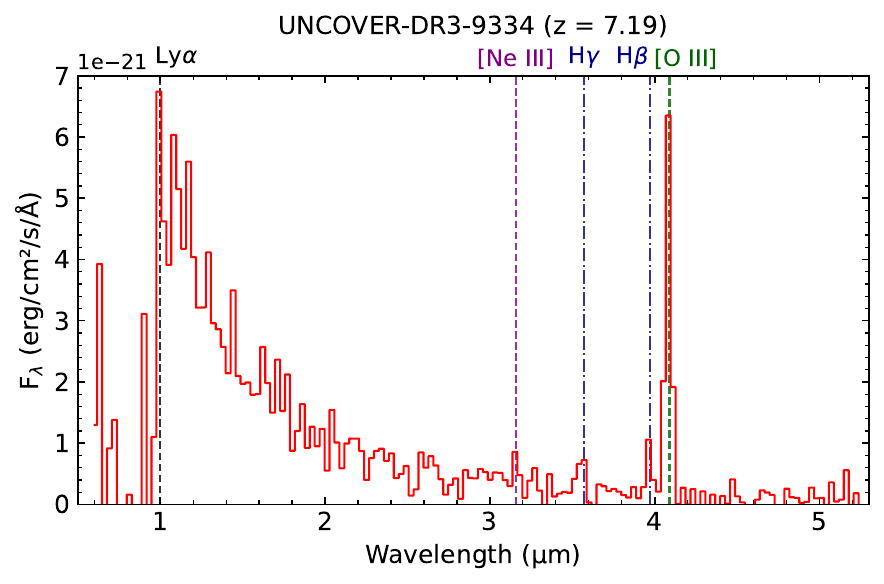} 
    \includegraphics[width=0.495\textwidth]{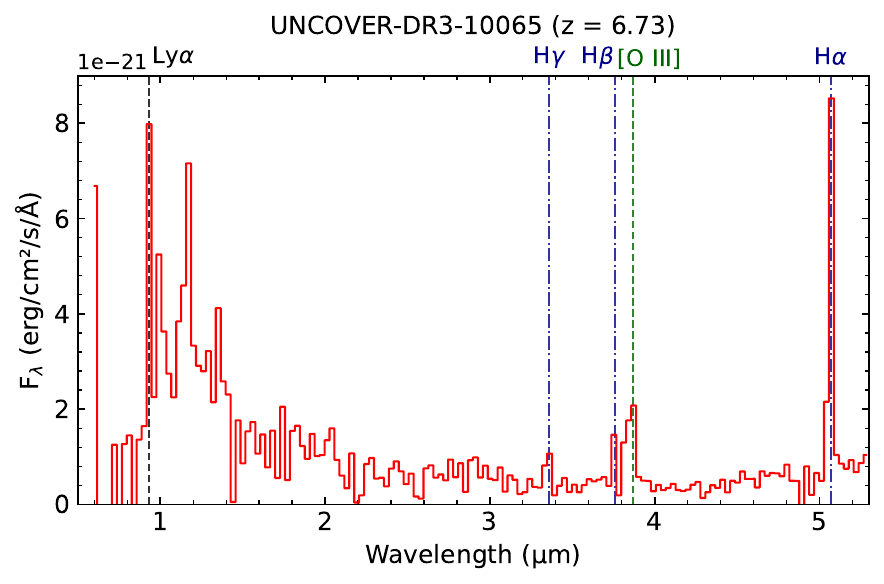} \\ 
    \includegraphics[width=0.495\textwidth]{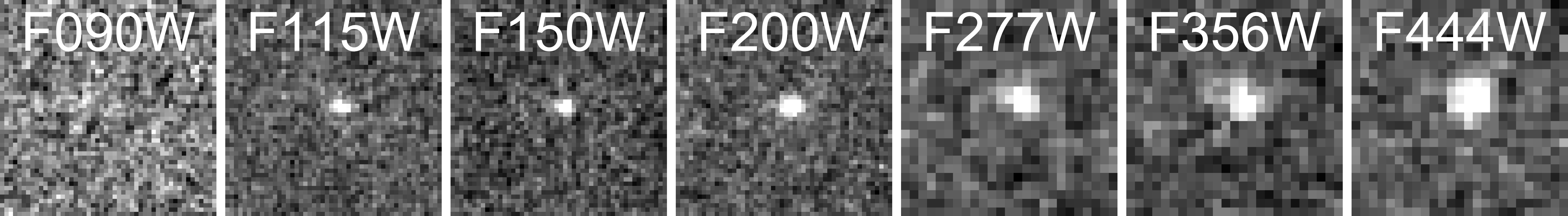} 
    \includegraphics[width=0.495\textwidth]{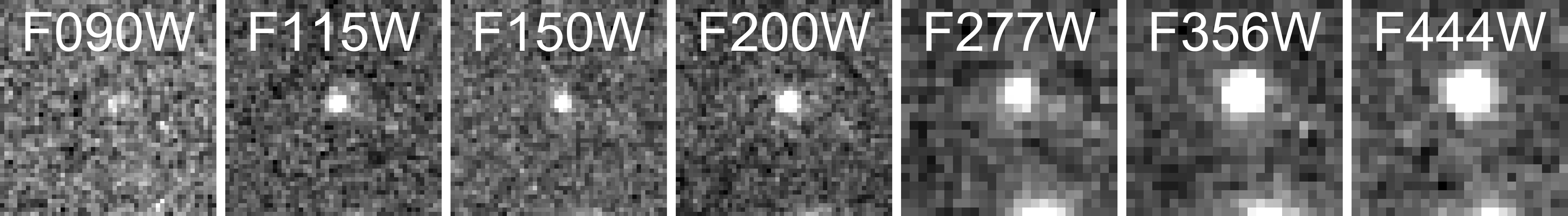} \\ 
    \includegraphics[width=0.495\textwidth]{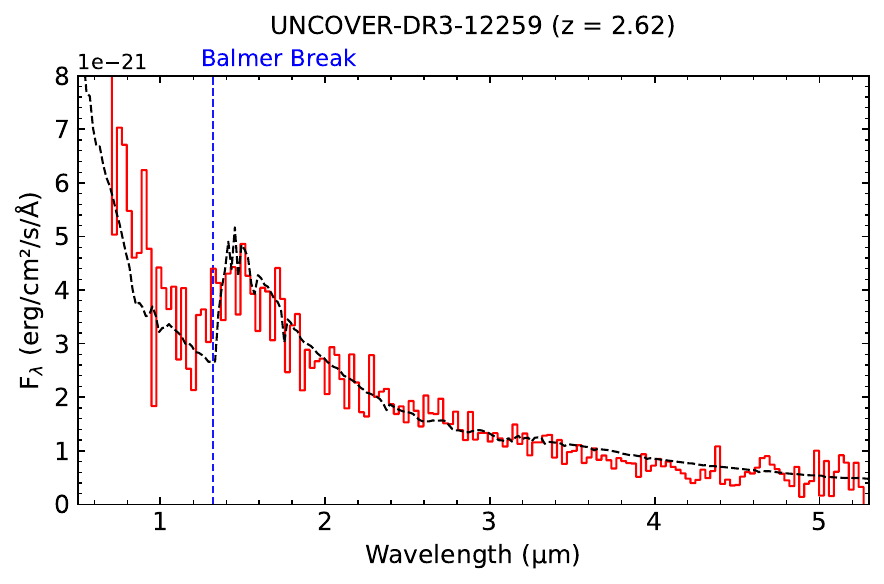} 
    \includegraphics[width=0.495\textwidth]{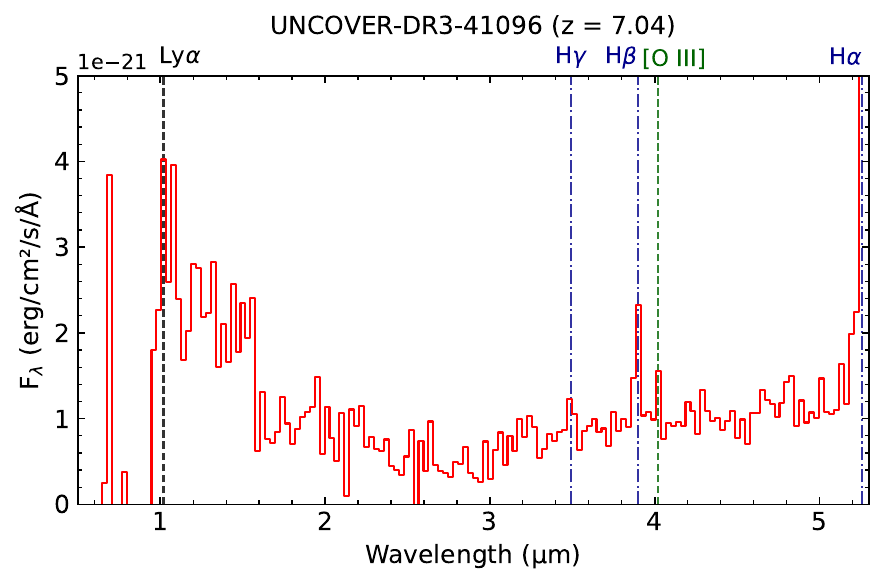} \\
    \includegraphics[width=0.495\textwidth]{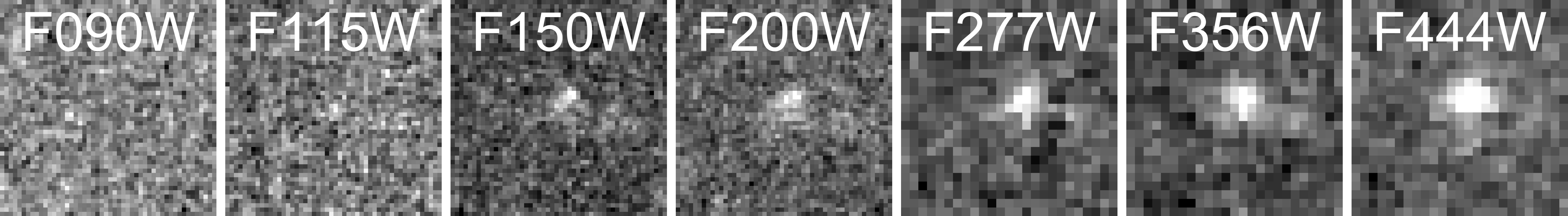} 
    \includegraphics[width=0.495\textwidth]{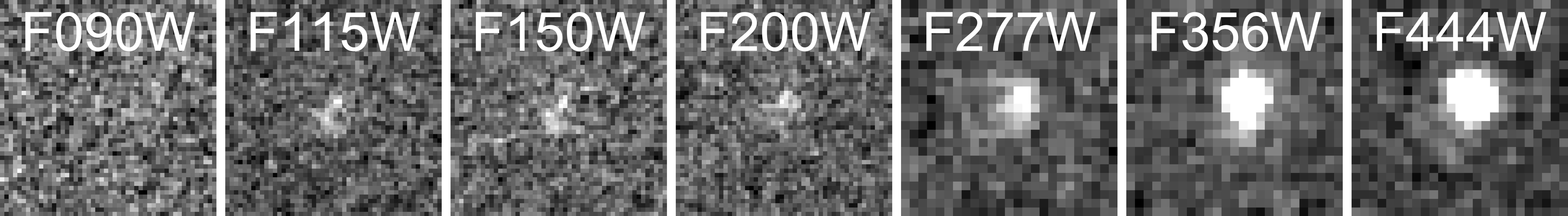} \\
    \caption{
    NIRSpec spectra of four high-redshift galaxies. Identified redshifted spectral features Ly$\alpha\,\lambda1216$, [Ne\,{\sc iii}]$\lambda3869$, H$\gamma$, H$\beta$, [O\,{\sc iii}]$\lambda5007$, H$\alpha$, and the Balmer break are indicated at the top of each panels. Their \textit{JWST} images ($1\arcsec$ on a side; north up, east to the left) are displayed at the bottom of each panel with filter names labelled.}
    \label{fgspec}
\end{figure*}

\begin{figure*}
    \includegraphics[width=0.495\textwidth]{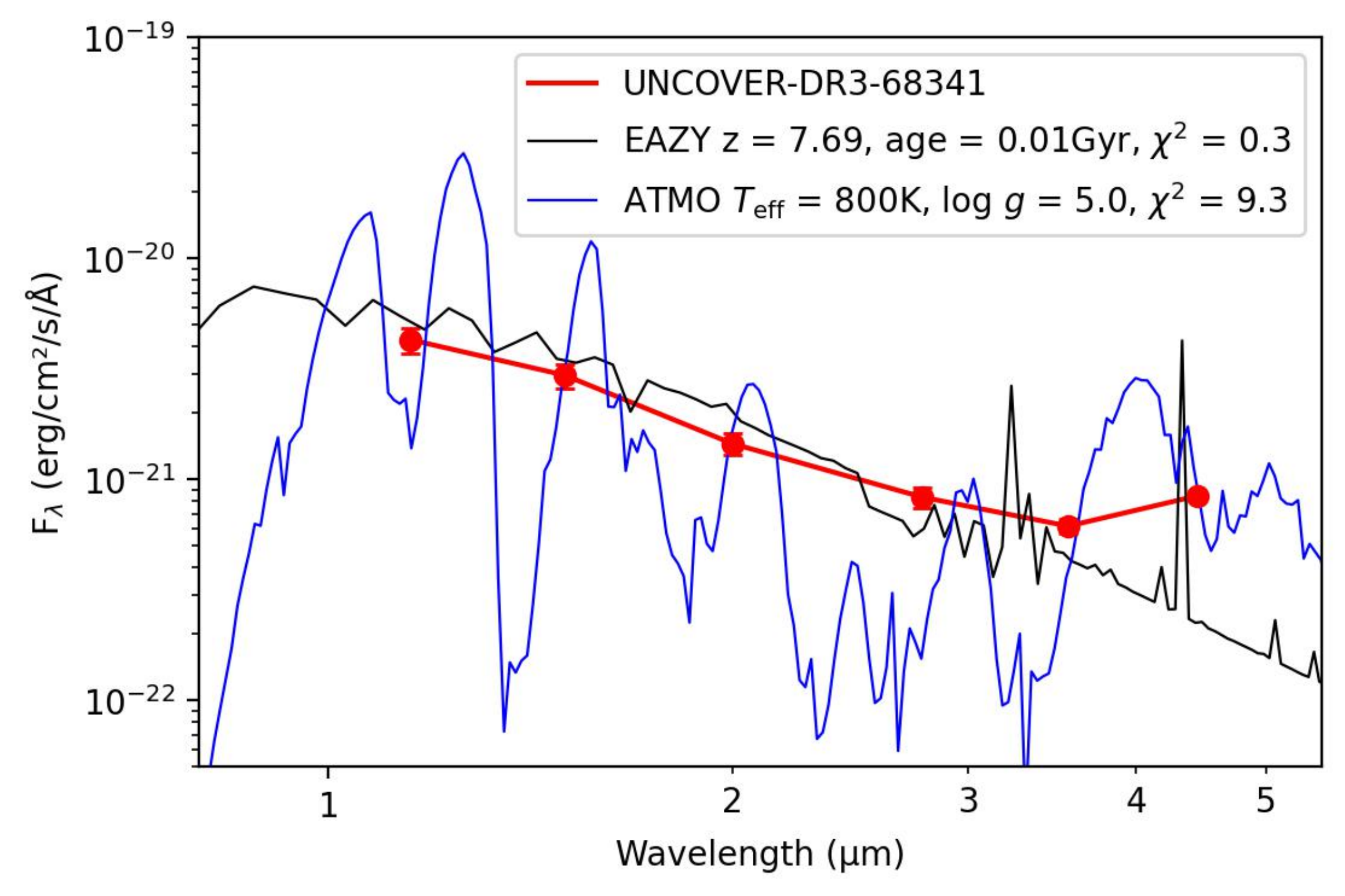} \\
    \includegraphics[width=0.495\textwidth]{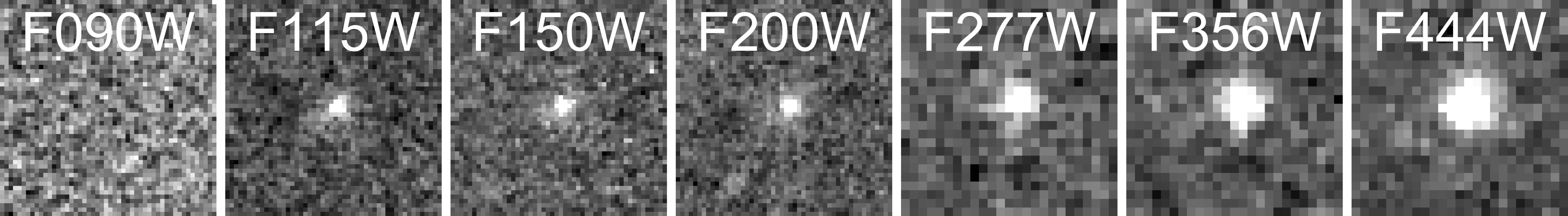} \\
    \includegraphics[width=0.495\textwidth]{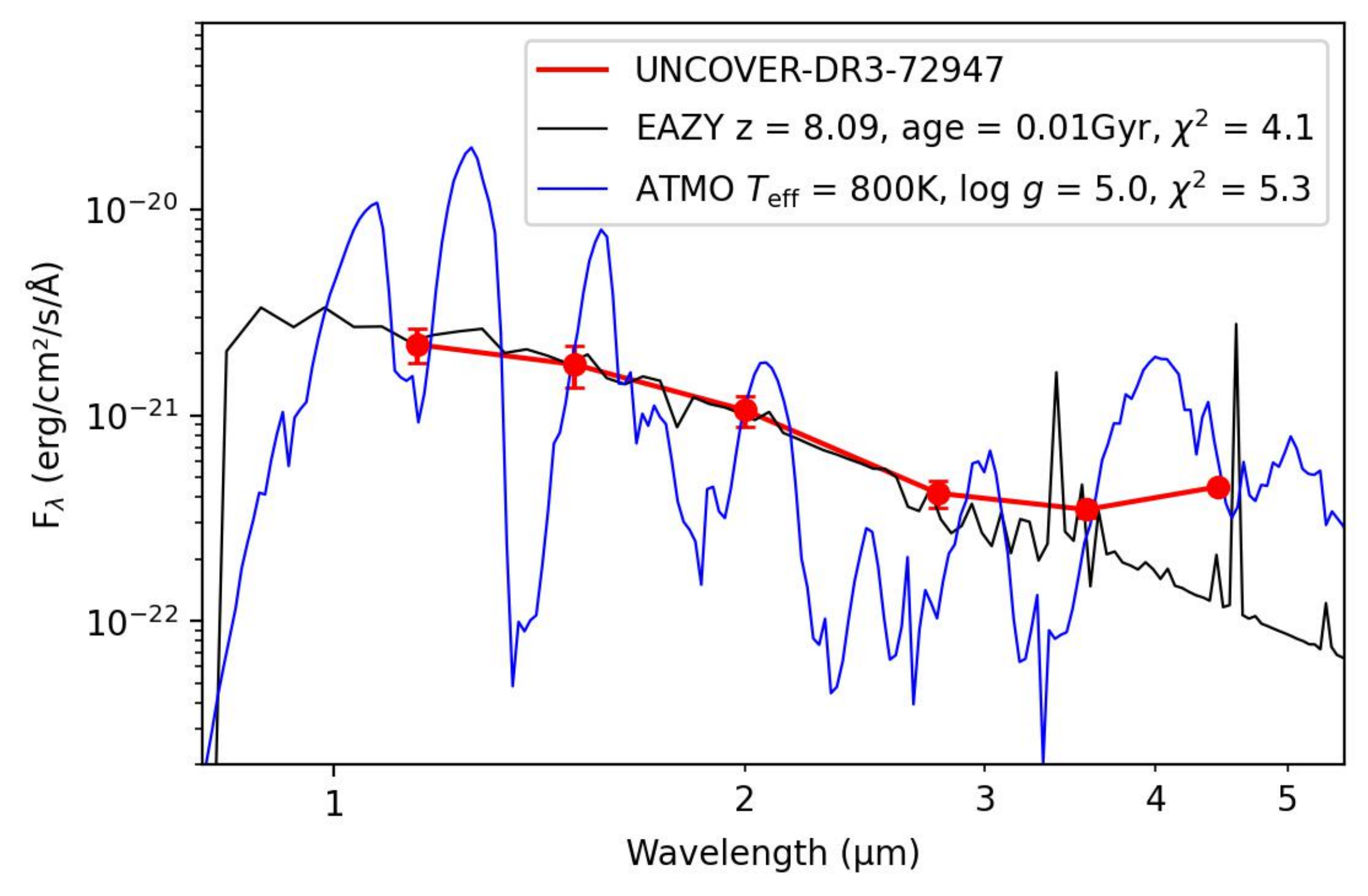} 
    \includegraphics[width=0.495\linewidth]{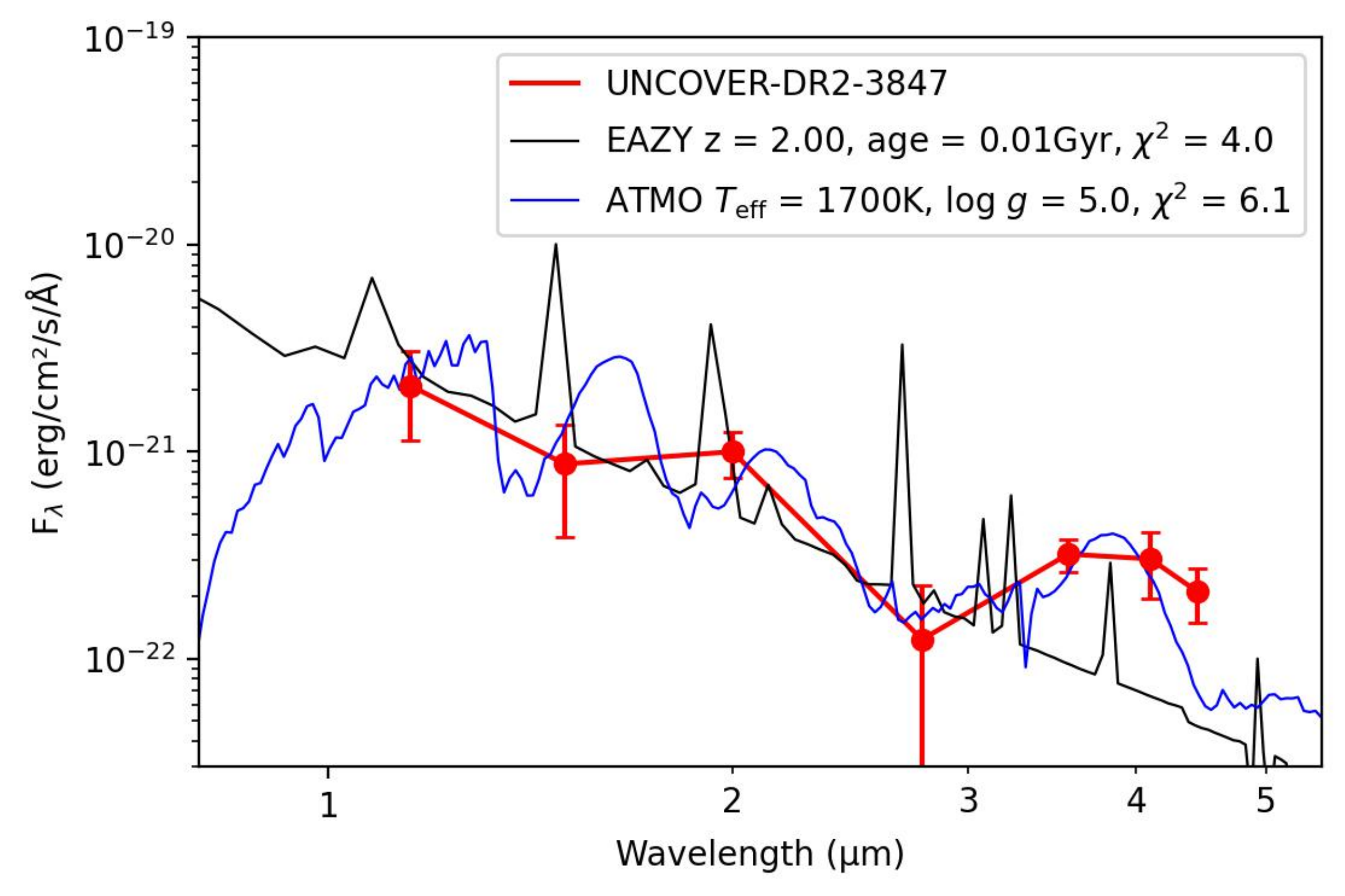} \\
    \includegraphics[width=0.495\textwidth]{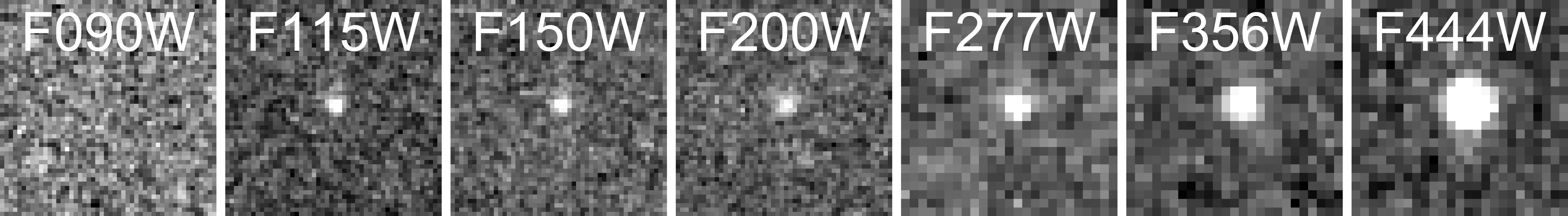} 
    \includegraphics[width=0.495\linewidth]{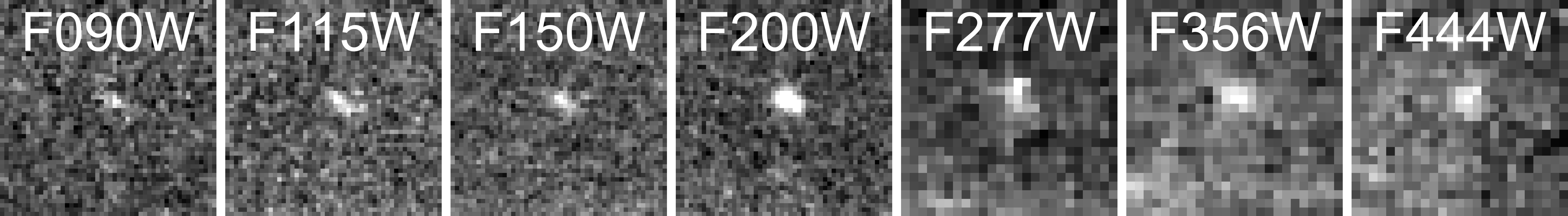} \\    \includegraphics[width=0.495\textwidth]{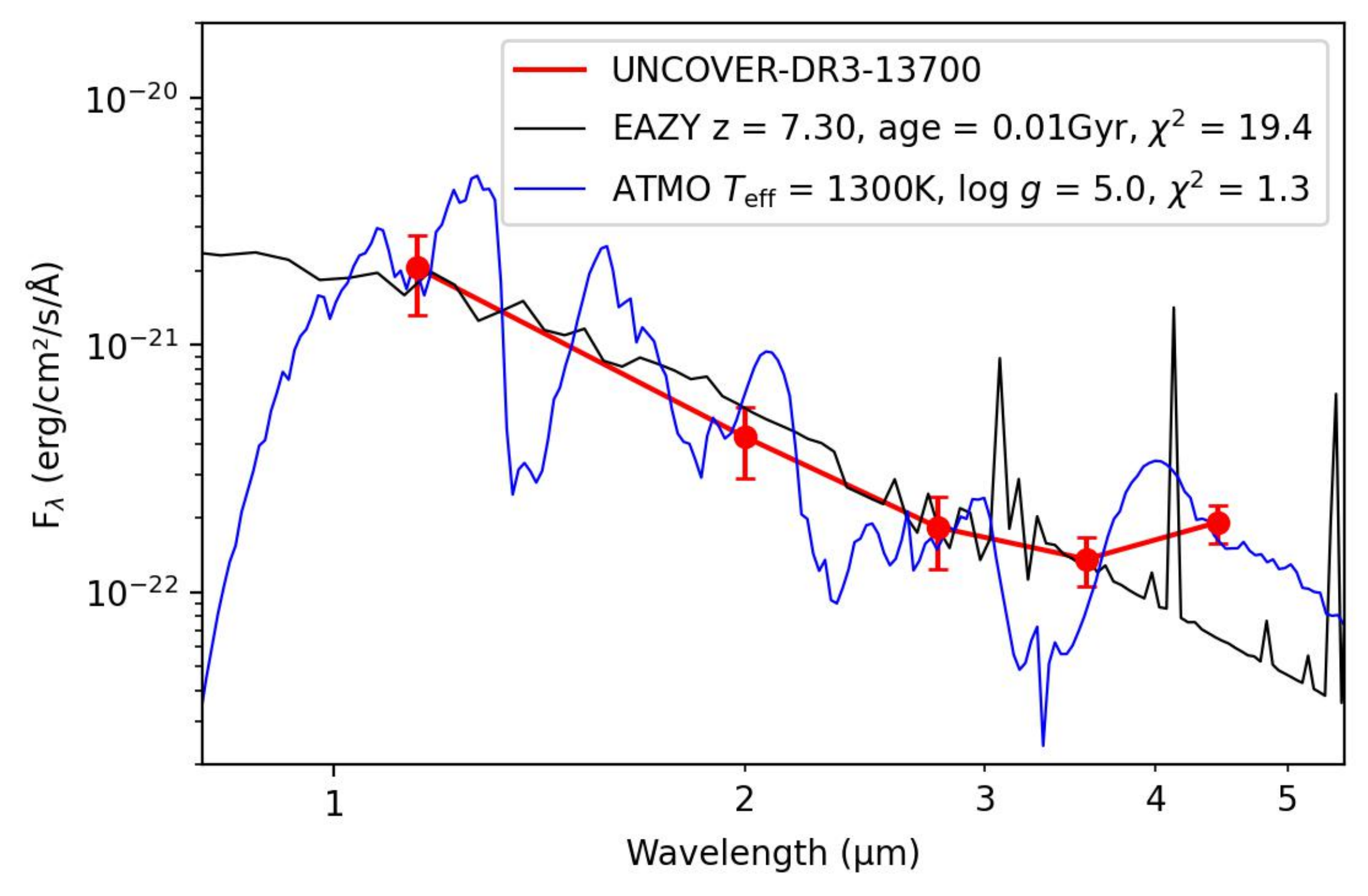}
    \includegraphics[width=0.495\textwidth]{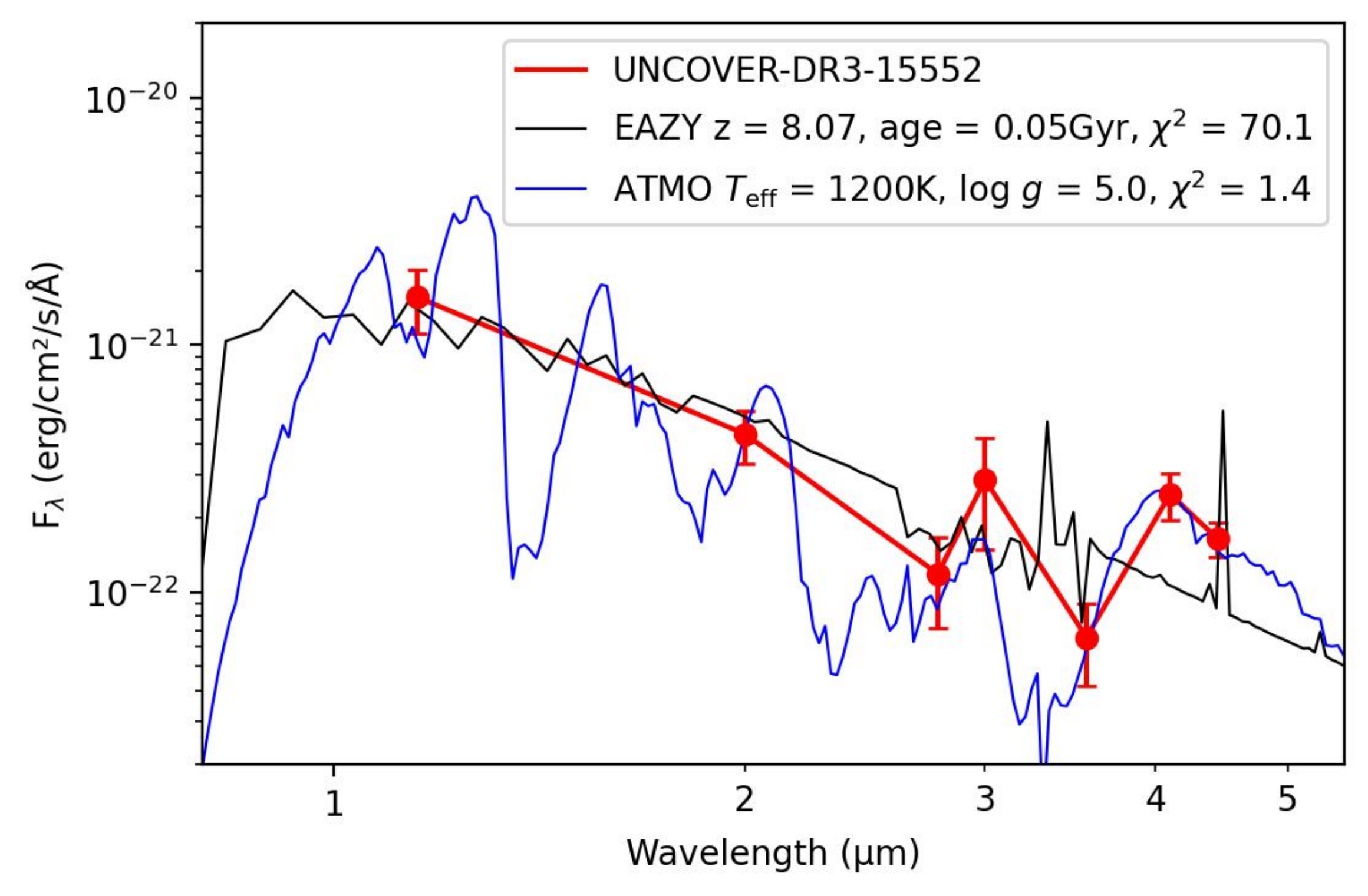} \\
    \includegraphics[width=0.495\textwidth]{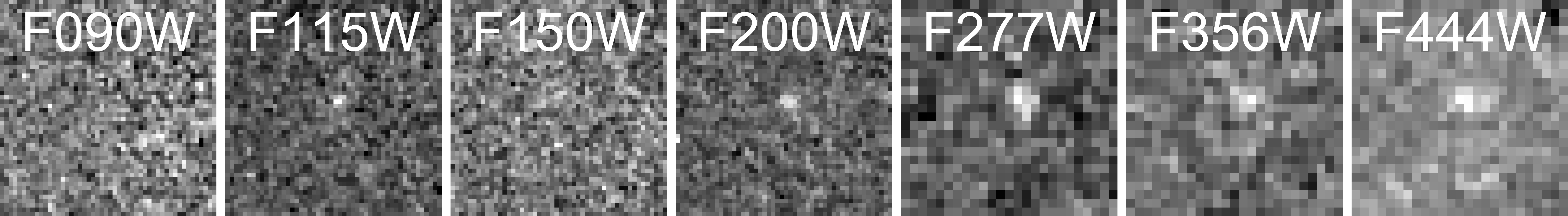}
    \includegraphics[width=0.495\textwidth]{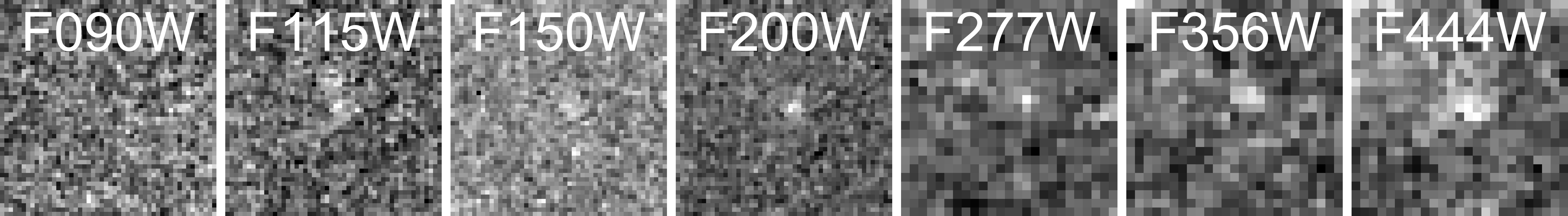} 
    \caption{The best-fitting galaxy and substellar models to the SEDs of one strong high-redshift galaxy candidates (top panel) and four ambiguous objects (middle and bottom panels). Their \textit{JWST} images ($1\arcsec$ on a side; north up, east to the left) are displayed at the bottom of each panel with filter names labelled.
    }
    \label{fgal}
\end{figure*}

Table~\ref{tgala} presents four spectroscopically confirmed high-redshift galaxies and one strong photometric candidate galaxy (UNCOVER-DR3-68341), as well as four ambiguous objects (UNCOVER-DR3-72947, UNCOVER-DR3-3847, UNCOVER-DR3-13700, and UNCOVER-DR3-15552) for which the photometry cannot reliably distinguish between high-redshift galaxies and BDs. Notably, the SEDs of the confirmed galaxies show better fits to substellar atmosphere models than to galaxy templates (Table~\ref{tgala}), highlighting the limitations of BD identification based solely on NIRCam photometry. The NIRSpec spectra of the four confirmed galaxies are displayed in Fig.~\ref{fgspec}, while the SEDs of the other five objects are shown in Fig.~\ref{fgal}.

%\begin{landscape}
\begin{table*}
    \caption{NIRCam photometry of four spectroscopically confirmed high-redshift galaxies (rows 1--4) and one strong photometric candidate galaxy (row 5), as well as four ambiguous objects (rows 6--9). The $\chi^2_{\rm V}$ is for best ATMO model in VOSA SED fitting. The $\chi^2_{\rm E}$ is for best  galaxy template in \textsc{eazy} SED fitting. }
    \label{tgala}
    \centering
    \begin{tabular}{lcc ccc ccc c}
    \hline\hline
         UNCOVER ID &  RA &  DEC & F115W & F150W &  F277W &  F444W & $\chi^2_{\rm V}$ & $\chi^2_{\rm E}$ & $z$\\
    \hline
        UNCOVER-DR3-9334 & 3.580667631 & $-$30.43479796 & 28.37 $\pm$ 0.19 & 28.32 $\pm$ 0.19 &  28.56 $\pm$ 0.13 &  27.45 $\pm$ 0.08 & 6.7 & 21.3 & 7.19\\ 
        UNCOVER-DR3-10065 & 3.592419869 & $-$30.43282806 & 28.52 $\pm$ 0.12 & 28.81 $\pm$ 0.18 &  28.68 $\pm$ 0.09 & 27.42 $\pm$ 0.04 & 18.0 & 39.5 & 6.73 \\ 
        UNCOVER-DR3-12259 & 3.571444921 & $-$30.42698725 & 29.80 $\pm$ 0.60 & 28.34 $\pm$ 0.16 &  28.44 $\pm$ 0.15 &  27.31 $\pm$ 0.08 & 5.6 & 22.3 & 2.62 \\ 
        UNCOVER-DR3-41096 & 3.569595059 & $-$30.37322117 & 28.44 $\pm$ 0.17 & 28.66 $\pm$ 0.17 &  28.87 $\pm$ 0.17 & 26.74 $\pm$ 0.03 & 24.4 & 31.3 & 7.04\\ 
        UNCOVER-DR3-68341 & 3.466752147 & $-$30.32039061 & 28.21 $\pm$ 0.13 & 28.03 $\pm$ 0.13 &  28.08 $\pm$ 0.11 &  27.04 $\pm$ 0.03 & 9.3 & 0.3 & 7.69 \\ 
        UNCOVER-DR3-72947 & 3.504801197 & $-$30.30183467 & 28.93 $\pm$ 0.19 & 28.60 $\pm$ 0.22 &  28.83 $\pm$ 0.15 &  27.73 $\pm$ 0.04 & 5.3 & 4.1 & 8.19\\ 
        UNCOVER-DR2-3847 & 3.603160115 & $-$30.42908969 & 28.99 $\pm$ 0.41 & 29.36 $\pm$ 0.48 & 30.15 $\pm$ 0.66 & 28.54 $\pm$ 0.28 & 6.1 & 4.0 & 2.00  \\
        UNCOVER-DR3-13700 & 3.616062090 & $-$30.42332913 &  29.01 $\pm$ 0.33 & 29.13 $\pm$ 0.56 &  29.72 $\pm$ 0.31 &  28.65 $\pm$ 0.17 & 1.3 & 19.4  & 7.30  \\ 
        UNCOVER-DR3-15552 & 3.608268145 & $-$30.41879402 & 29.30 $\pm$ 0.28 & 31.62 $\pm$ 1.59 &  30.20 $\pm$ 0.36 &   28.81 $\pm$ 0.16 & 1.4 & 70.1 & 8.07 \\   
    \hline
    \end{tabular} \\
\end{table*}

%%%%%%%%%%%%%%%%%%%%%%%%%%%%%%%%%%%%%%%%%%%%%%%%%%

% Don't change these lines
\bsp	% typesetting comment
\label{lastpage}
\end{document}